\documentclass[prd,aps,twocolumn,amsmath,amssymb,floatfix,nofootinbib,superscriptaddress]{revtex4}

\usepackage{t1enc}
\usepackage{amsfonts}
\usepackage{gensymb}
\usepackage{graphicx}
\usepackage{float}
\usepackage{physics}
\usepackage{booktabs}

\newcommand{\drm}{\ensuremath{\mathrm{d}}}
\newcommand{\dcal}{\ensuremath{\mathcal{D}}}
\newcommand{\Seff}{\ensuremath{S_{\mathrm{eff}}}}
\newcommand{\Seffo}{\ensuremath{S_{\mathrm{eff,0}}}}
\newcommand{\Seffi}{\ensuremath{S_{\mathrm{eff,1}}}}

\begin{document}

\title{Exponential reduction of the sign problem at finite density\\
in the 2+1D XY model via contour deformations}

\author{Matteo Giordano}
\affiliation{ELTE E\"otv\"os Lor\'and University, Institute for Theoretical Physics, P\'azm\'any P\'eter s\'et\'any 1/A, H-1117, Budapest, Hungary}

\author{Korn\'el Kap\'as}
\affiliation{ELTE E\"otv\"os Lor\'and University, Institute for Theoretical Physics, P\'azm\'any P\'eter s\'et\'any 1/A, H-1117, Budapest, Hungary}

\author{S\'andor D. Katz}
\affiliation{ELTE E\"otv\"os Lor\'and University, Institute for Theoretical Physics, P\'azm\'any P\'eter s\'et\'any 1/A, H-1117, Budapest, Hungary}
\affiliation{MTA-ELTE Theoretical Physics Research Group, P\'azm\'any P\'eter s\'et\'any 1/A, H-1117 Budapest, Hungary.}

\author{Attila P\'asztor}
\affiliation{ELTE E\"otv\"os Lor\'and University, Institute for Theoretical Physics, P\'azm\'any P\'eter s\'et\'any 1/A, H-1117, Budapest, Hungary}

\author{Zolt\'an Tulip\'ant}
\affiliation{ELTE E\"otv\"os Lor\'and University, Institute for Theoretical Physics, P\'azm\'any P\'eter s\'et\'any 1/A, H-1117, Budapest, Hungary}
\begin{abstract}
We study the 2+1 dimensional XY model at nonzero chemical 
potential $\mu$ on deformed integration manifolds, with the aim 
of alleviating 
its sign problem. We investigate several proposals for the 
deformations, and considerably improve on the severity of 
the sign problem with respect to standard reweighting approaches. 
We present numerical evidence that the reduction of the 
sign problem is exponential both in $\mu^2$ and in the spatial
volume. We also present a new approach to the 
optimization procedure based on reweighting, that 
sensibly reduces its computational cost.
\end{abstract}

\maketitle

\section{Introduction}

Euclidean quantum field theories with a finite chemical 
potential generally suffer from a complex action problem: 
the path integral weights are complex, and therefore cannot be interpreted as the Boltzmann weights of a classical statistical mechanical system. In QCD, this complex action problem is a severe roadblock for first principles understanding of the physics of neutron stars, supernovae, as well as heavy ion collisions at lower collision energies. 
In some theories the sign problem can be solved by a reformulation 
of the theory in different variables, such that in the new variables
the weights are manifestly real and 
positive~\cite{Chandrasekharan:1999cm,Alford:2001ug,Endres:2006xu,Bruckmann:2015sua,Gattringer:2015nea}. 
This has not been achieved in QCD so far. 

The existence of a sign problem does not make simulations completely impossible.
In the presence of a complex action problem, simulations can still be carried out by standard Monte Carlo methods in the phase-quenched (PQ) theory, with Boltzmann weights proportional to $\left| e^{-S} \right|$, or - assuming that the grandcanonical partition function is real - the sign-quenched (SQ) theory, with weights proportional to $\left| \operatorname{Re} e^{-S} \right|$.
After such simulations have been carried out, the ratio of the simulated and target partition functions, as well as the expectation values of different operators $O$, can in principle be reconstructed via the formulas:
\begin{equation}
    \label{eq:reweighting}
    \begin{aligned}
        \frac{Z}{Z_{\textrm{PQ}}} &= \left\langle \cos \theta \right\rangle_{\textrm{PQ}} \textrm{,}\quad\quad
        \frac{Z}{Z_{\textrm{SQ}}} = \left\langle \frac{\cos\theta}{\left|\cos\theta\right|}\right\rangle_{\textrm{SQ}} \textrm{,} \\
\left\langle O\right\rangle &= \frac{\left\langle O e^{i\theta}\right\rangle_{\textrm{PQ}}}{\left\langle e^{i\theta} \right\rangle_{\textrm{PQ}}} = \frac{\left\langle O \frac{\cos\theta}{\left|\cos\theta\right|}\right\rangle_{\textrm{SQ}}}{\left\langle\frac{\cos\theta}{\left|\cos\theta\right|}\right\rangle_{\textrm{SQ}}} \rm{,}
    \end{aligned}
\end{equation}
where we introduced the phase of the complex action $\theta$ as $e^{-S}=\left| e^{-S} \right| e^{i \theta}$. This leads to large cancellations when the phases $e^{i \theta}$ have large fluctuations, a problem that is generally referred to as a sign problem. 
The severity of the sign problem can be measured by the partition function ratios $Z/Z_{\textrm{PQ}}$ and $Z/Z_{\textrm{SQ}}$,
and as long as these quantities are under numerical control 
one can reconstruct expectation values in the desired 
theory reliably. In fact, reweighting 
from the phase and sign quenched theories is starting to become 
feasible in QCD~\cite{Fodor:2007vv,Endrodi:2018zda,Giordano:2020roi,Borsanyi:2021hbk}.
However, the range of applicability - both in the chemical potential and in the physical volume - of such an approach is
severely constrained by the sign problem, 
which requires an exponential increase of the statistics both as a function of the volume $V$ and of the chemical potential $\mu$. It is therefore desirable to develop methods that either 
solve or at least alleviate the sign problem. 
Even if only the second goal is achieved, this could still drastically 
increase the range of parameters that reweighting methods can reach. 

A set of methods that try to deal with the sign problem is based on complexification of the fields. There are two broadly defined approaches of this type. In the first one - complex Langevin~\cite{Parisi:1983mgm, Aarts:2009uq,Seiler:2012wz, Aarts:2017vrv, Scherzer:2018hid} - an $N$-dimensional integration over real fields is enlarged to a $2N$-dimensional integral over the real components of the complexified fields. In the second one - contour deformations - the integral still remains $N$-dimensional, but the integration manifold is deformed to a different manifold of the same dimension. 
In this paper, we pursue this second approach.

In most cases of interest, the path integral weights are holomophic functions of the 
fields.\footnote{A notable exception is lattice QCD with rooted staggered 
fermions~\cite{Golterman:2006rw,Giordano:2019gev}.} In such a case, any integration 
manifold that is in the same homology class as the undeformed manifold leads to the same 
partition function~\cite{Alexandru:2020wrj}. However, the phase and sign quenched integrands are not holomorphic, and therefore 
the phase and sign quenched partition functions are not invariant under such contour deformations. Thus, it could be possible to bring closer to unity the ratios $Z/Z_{\textrm{PQ}}$ and $Z/Z_\textrm{SQ}$, measuring the severity of the sign problem, by such contour deformations. One could then perform simulations in the contour-deformed phase or
sign quenched theory, and perform a reweighting via Eq.~\eqref{eq:reweighting} to get results in the 
target theory.

There are different ways to deform integration contours to make sign problems milder. First, there are methods based on Lefschetz thimbles~\cite{Cristoforetti:2012su, Cristoforetti:2013wha, Alexandru:2015sua, Fukuma:2020fez, Alexandru:2020wrj}. Second, there are the more ad hoc path optimization methods which we pursue here. 
The main idea of these methods is to parametrize the deformed integration manifold by some finite number of parameters, which are then adjusted to make the sign problem as mild as possible. In the context of the sign problem at nonzero chemical potential, the path optimization method was applied to a one-dimensional oscillating integral~\cite{Mori:2017pne}, the 0+1D $\phi^4$ theory~\cite{Bursa:2018ykf}, the 0+1D PNJL model~\cite{Kashiwa:2018vxr}, 0+1D QCD~\cite{Mori:2019tux}, the 1+1D $\phi^4$ model~\cite{Mori:2017nwj}, the 1+1D Thirring model~\cite{Alexandru:2018fqp}, the 2+1D Thirring model~\cite{Alexandru:2018ddf}, and Bose gases of several dimensions~\cite{Bursa:2021org}. Other applications of the sign optimization method include improving the signal-to-noise ratio of noisy observables at zero chemical potential in 0+1D scalar field theory and 1+1D U(1) gauge theory~\cite{Detmold:2020ncp} and in 1+1D SU(2) and SU(3) gauge theory~\cite{Detmold:2021ulb}, and reduction of the sign problem in 1+1D U(1) gauge theory with complex coupling constants~\cite{Kashiwa:2020brj}. 

In this paper we apply the path optimization method to the 2+1 dimensional XY model.
The choice of the model is mainly motivated by the fact that it shares several technical features with QCD, so that insight 
obtained here can hopefully be applied there as well.
First, the integration variables take values in a compact space, which
requires a slightly different treatment of contour deformations than
non-compact integration domains do. Second, this theory has
Roberge-Weiss periodicity at imaginary chemical
potential~\cite{Roberge:1986mm, Aarts:2009dg}. Third, the model also
shares the property of QCD that the complex Langevin approach fails
for small coupling
$\beta$~\cite{Aarts:2010aq,Scherzer:2019lrh}. Fourth, like in lattice
QCD, the dependence of the action on the chemical potential is
non-linear.

Finally, like in QCD, the effects of a chemical potential should
saturate at large $\mu$, where the temporal dependence of the relevant
field configurations tends to become trivial and $\mu$ essentially
drops out of expectation values; in this limit the sign problem
becomes mild.

Differently from QCD, however, the XY model 
can be rewritten in 
the worldline formulation to be free of a sign problem~\cite{Endres:2006xu,Banerjee:2010kc,Langfeld:2013kno}, allowing 
for direct simulations using a worm algorithm, and 
making explicit comparisons possible with 
the path optimization method.

Our goals in this work are the following. First, we investigate how much improvement can be 
made to the severity of the sign problem in this model by 
very simple ans\"atze for the contour deformations. Second, we study the chemical potential and volume dependence of the 
improvement on the sign problem, and so on the  statistics required for reliable reweighting, achieved with the optimized contours.
Third, we wish to see whether the optimization procedure can be performed more efficiently, avoiding a full Monte Carlo simulation at each step 
of the iteration searching for the optimum, by simply reweighting to the different contours using the same fixed ensembles.

The results for all three of these inquiries turn out to be quite encouraging. First - as we will see - a rather drastic improvement 
can be achieved in the severity of the sign problem, even with relatively simple ans\"atze. Second, we present numerical evidence
that the improvement is exponential, i.e., it reduces the exponent of the severity of the sign problem. And third, the optimization 
of the contour is feasible with reweighting only. 
This leads to a radical 
improvement in the cost of the optimization procedure itself.

The plan of the paper is the following. 
In Section~\ref{sect:model} we briefly  discuss the contour deformation approach to the 2+1 dimensional XY model at finite chemical potential. 
In Section~\ref{sect:opt} we provide details on the optimization procedure and on the different parametrizations used to ameliorate the severity of 
the sign problem. 
In Section~\ref{sect:results} we illustrate the chemical potential and volume dependence of the achieved improvement.
In Section~\ref{sect:rew} we present a modified optimization 
procedure based on
reweighting to different contours from a fixed 
ensemble, that allows us to reduce computational costs.
We summarize our conclusions in Section~\ref{sect:disc}. 

\section{Contour deformations for 2+1D XY model at nonzero chemical potential}
\label{sect:model}

The action of the 2+1D XY model with nonzero chemical potential \cite{Aarts:2010aq,Banerjee:2010kc,Langfeld:2013kno} is 
\begin{equation}
\label{O2action}
S=-\beta\sum_x\sum_{n=0}^2\cos(\varphi_x-\varphi_{x+\hat{n}}+\imath\mu\delta_{n0}),
\end{equation}
where the sum runs over all lattice sites and directions, with $0$ identified as the temporal direction. Periodic boundary conditions are imposed in every direction. The partition function, 
\begin{align}
Z(\mu) &= \int_{-\pi}^\pi\drm\varphi_{(000)} \dots \int_{-\pi}^\pi\drm\varphi_{(N_0N_1N_2)}\,e^{-S} \nonumber \\
&\equiv \int_{\mathcal{M}_0} \dcal\varphi\,e^{-S}
\end{align}
can be interpreted as a complex contour integral in each $\varphi_x$ 
with endpoints at $-\pi$ and $\pi$, and so as an integral over an $N_0 N_1 N_2$-dimensional manifold ${\mathcal{M}_0}={[-\pi,\pi]^{N_0 N_1 N_2}}$ embedded in a $2N_0 N_1 N_2$ dimensional space. 

\begin{figure}[t!]
    \centering
    \includegraphics[width=0.48\textwidth]{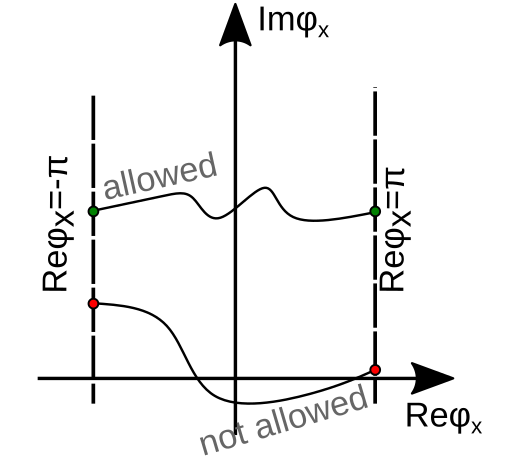}
    \caption{Illustration of an allowed and not allowed contour deformation for a single field variable.}
    \label{fig:illustrate}
\end{figure}

For each $\varphi_x$, we will consider contours in the strip of the 
complex plane satisfying $-\pi \leq \operatorname{Re} \varphi_x \leq \pi$, with the points
$-\pi + i\operatorname{Im} \varphi_x$ and $\pi + i\operatorname{Im} \varphi_x$ identified, i.e., a cylinder. As long as the 
contours remain smooth on this space, i.e., if they are smooth curves from $-\pi+ix$ to $\pi+ix$ for some $x \in \mathbb{R}$, the partition function remains unchanged.
This can be shown by connecting the original and the shifted contours using line 
segments perpendicular to the real axis, whose contributions cancel out thanks to 
periodicity of the integrand in $\operatorname{Re} \varphi_x$. Allowed contour deformations for a 
single field variable are illustrated in Fig.~\ref{fig:illustrate}.

Denoting the deformed manifold ${\cal M}$, and parametrizing it with real parameters $t_x$, 
we have for any ${\cal M}$ satisfying the requirements above
\begin{equation}
    \begin{aligned}
Z(\mu) &= \int_{{\mathcal{M}_0}} \dcal \varphi e^{-S} = \int_{\mathcal{M}} \dcal \varphi e^{-S} \\
       &= \int\dcal t \det J \,e^{-S} \equiv \int\dcal t\,e^{-\Seff},
    \end{aligned}
\end{equation}
where $J$ is the Jacobian matrix, with elements
\begin{equation}
    J_{xy} = \frac{\partial \varphi_x}{\partial t_y},
\end{equation}
and the effective action is defined as $\Seff=S-\ln\det J$. Exploiting the reality of the partition function we can write it in a manifestly real form, 
\begin{equation}
Z(\mu) = \int\dcal t\,\cos\Seff^I\,e^{-\Seff^R}, 
\end{equation} where $\Seff^R$ and $\Seff^I$ are the real and imaginary parts of the effective action. This enables us to make use of the sign reweighting approach~\cite{deForcrand:2002pa,Alexandru:2005ix,Giordano:2020roi,Borsanyi:2021hbk} by using the absolute value of the integrand, $|\cos\Seff^I|\,e^{-\Seff^R}\geq0$, as a weight in importance sampling. This method has a slightly milder sign problem than the phase reweighting method~\cite{deForcrand:2002pa,Borsanyi:2021hbk}.
The corresponding expectation values will be denoted as $\langle\dots\rangle_{\mathrm{SQ},\mu}$. The severity of the sign problem is measured by the average sign, 
\begin{equation}
\langle\varepsilon\rangle_{\mathrm{SQ},\mu} = 
\frac{\int\dcal t\,\cos\Seff^I\,e^{-\Seff^R}}{\int\dcal t\,|\cos\Seff^I|\,e^{-\Seff^R}}.
\end{equation}
While the numerator of this expression is invariant under changing integration contours, the denominator is 
instead altered. Hence, deforming contours leaves the target partition function unchanged but it has an effect on the 
severity of the sign problem. 

\section{Optimization of the integration manifold}
\label{sect:opt}

In this paper we work at $\beta=0.4$, that for $\mu=0$ is in the
disordered phase of the
model~\cite{Aarts:2010aq,Banerjee:2010kc,Langfeld:2013kno}. 
As a preliminary check, we performed a scan in $\mu$ using the sign
problem-free worldline formulation and the worm algorithm of
Ref.~\cite{Banerjee:2010kc}. Using a standard finite size scaling
analysis, we observe a phase transition at $\mu_c^2 \approx 0.54$.

Our goal is to find integration contours that give a larger $\langle\varepsilon\rangle_{\mathrm{SQ},\mu}$ than without contour deformation. As such, we need to maximize the expectation value of the sign, or alternatively, minimize a cost function, with respect to some coefficients $p_i$ 
that parametrize the contours. As our cost function, we choose the ratio of the number of configurations with negative and positive $\cos\Seff^I$, 
\begin{equation}
\frac{N_-}{N_+} =
\frac{\int\dcal t\,\Theta(-\cos\Seff^I)|\cos\Seff^I|\,e^{-\Seff^R}}
{\int\dcal t\,\Theta(\cos\Seff^I)|\cos\Seff^I|\,e^{-\Seff^R}} \textrm{.}
\end{equation}
Its gradient can be computed easily, 
\begin{equation}
\mathcal{F}_i \equiv
\frac{\partial}{\partial p_i}\left(\frac{N_-}{N_+}\right) = \frac{N_-}{N_+}\Big(\langle F_i\rangle_- - \langle F_i\rangle_+\Big),
\end{equation}
where $\langle\dots\rangle_{\pm}$ means averaging only over configurations with positive/negative $\cos \Seff^I$, and 
\begin{equation}
F_i \equiv -\frac{\partial\Seff^R}{\partial p_i} - \tan\Seff^I\frac{\partial\Seff^I}{\partial p_i}\rm.
\end{equation}
We perform optimization using a simple gradient descent algorithm. At every optimization step we update coefficients by subtracting the gradient of the cost function with a multiplication factor, 
\begin{equation}
\boldsymbol{p}^{(j+1)} = \boldsymbol{p}^{(j)} - \alpha_j \boldsymbol{\mathcal{F}}^{(j)},
\end{equation}
where $\boldsymbol{p}^{(j)}$ is the vector of coefficients obtained at the $j$th step, $\boldsymbol{\mathcal{F}}^{(j)}$ is 
the value of the gradient obtained using $\boldsymbol{p}^{(j)}$, and $\alpha_j$ is 
\begin{equation}
\alpha _j = \frac{\left| \left(\boldsymbol{p}^{(j)}-\boldsymbol{p}^{(j-1)}\right)^T\cdot \left(\boldsymbol{\mathcal{F}}^{(j)}-\boldsymbol{\mathcal{F}}^{(j-1)}\right) \right|}{||\boldsymbol{\mathcal{F}}^{(j)}-\boldsymbol{\mathcal{F}}^{(j-1)} ||^2}.
\end{equation}
The algorithm terminates when the relative change of the coefficients decreases below a prescribed tolerance value.

We note that since we always start the optimization near the undeformed integration manifold, our procedure explores only the vicinity of the original contour, and could not detect an optimum lying beyond a ``potential barrier''.

We parametrize the integration manifold by the real parts of the field variables on the different lattice sites, denoted by $t_x$. 
Periodicity of the action in the real part of $\varphi_x$ restricts
the imaginary part of $\varphi_x$ on the deformed manifold to be a
Fourier series in $t_x$, with coefficients that in general depend on
$x$ and on the other field variables. As our first attempt, we
parametrized each complex $\varphi_x$ using only the corresponding
$t_x$.  However, the sign problem could not be improved this way, as
using this parametrization the original contour turned out to be the
local minimum of the cost function. In retrospect, this is not surprising, since any
nontrivial such parametrization breaks the symmetry of the model under
a common shift $t_x \to t_x + c$. This suggests to restrict to a
Fourier series in sines and cosines of $t_x-t_y$.  We then
experimented with different ans\"atze that couple neighbouring sites in
simple ways, all of which could improve the sign problem, but by quite different amounts.

\subsection{Ans\"atze without temporal translational invariance}

\begin{figure}[b!]
    \centering
    \includegraphics[width=0.48\textwidth]{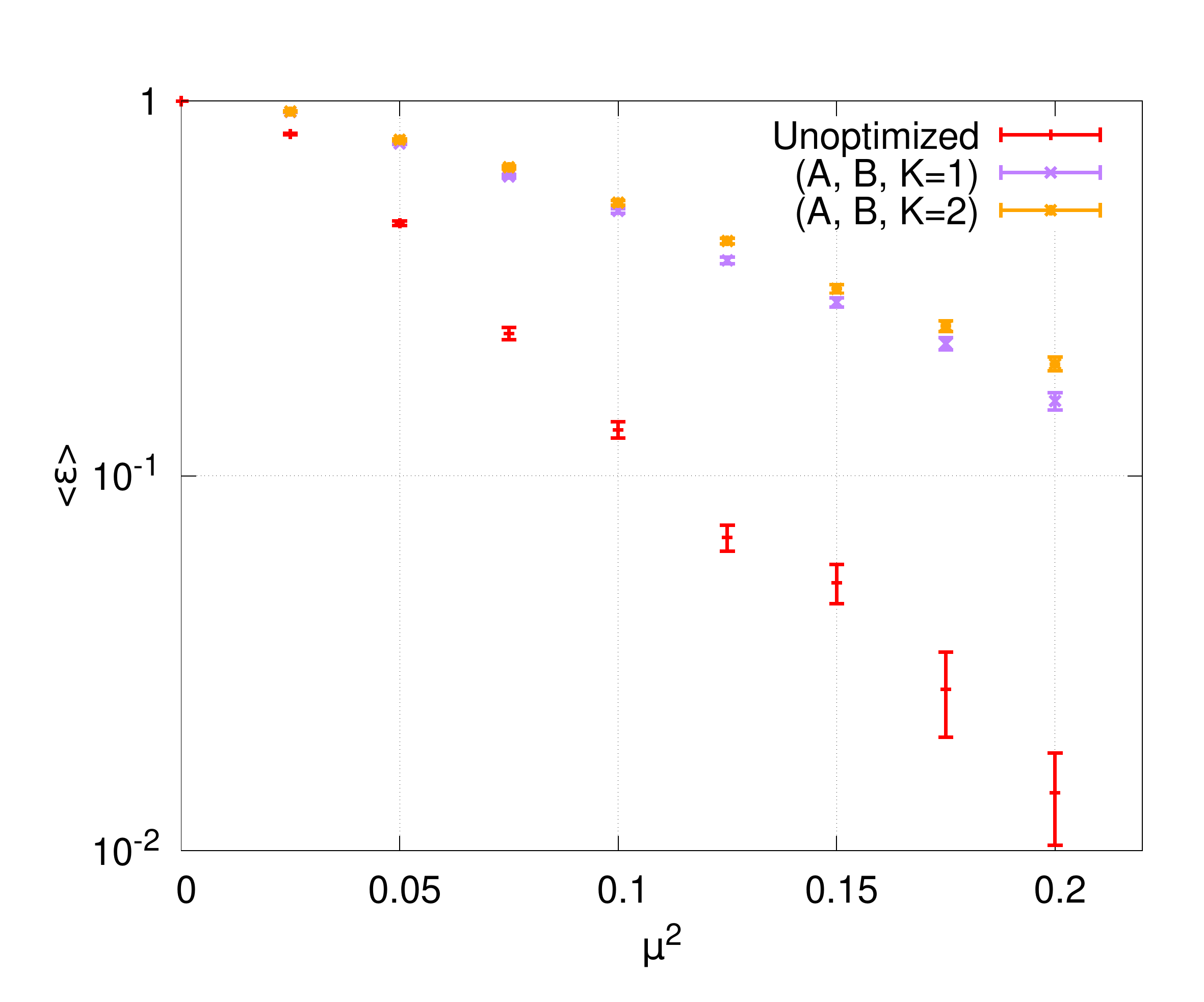}
    \caption{Severity of the sign problem on an $\Omega=8^3$ lattice with $\beta=0.4$. Unimproved results are shown in red and the results optimized obtained using Eq.~\eqref{fullparam} for $K=1$ and $2$ are shown in purple and orange, respectively.}
    \label{fig:3Dsign}
\end{figure}

\begin{figure}[t!]
    \centering
    \includegraphics[width=0.48\textwidth]{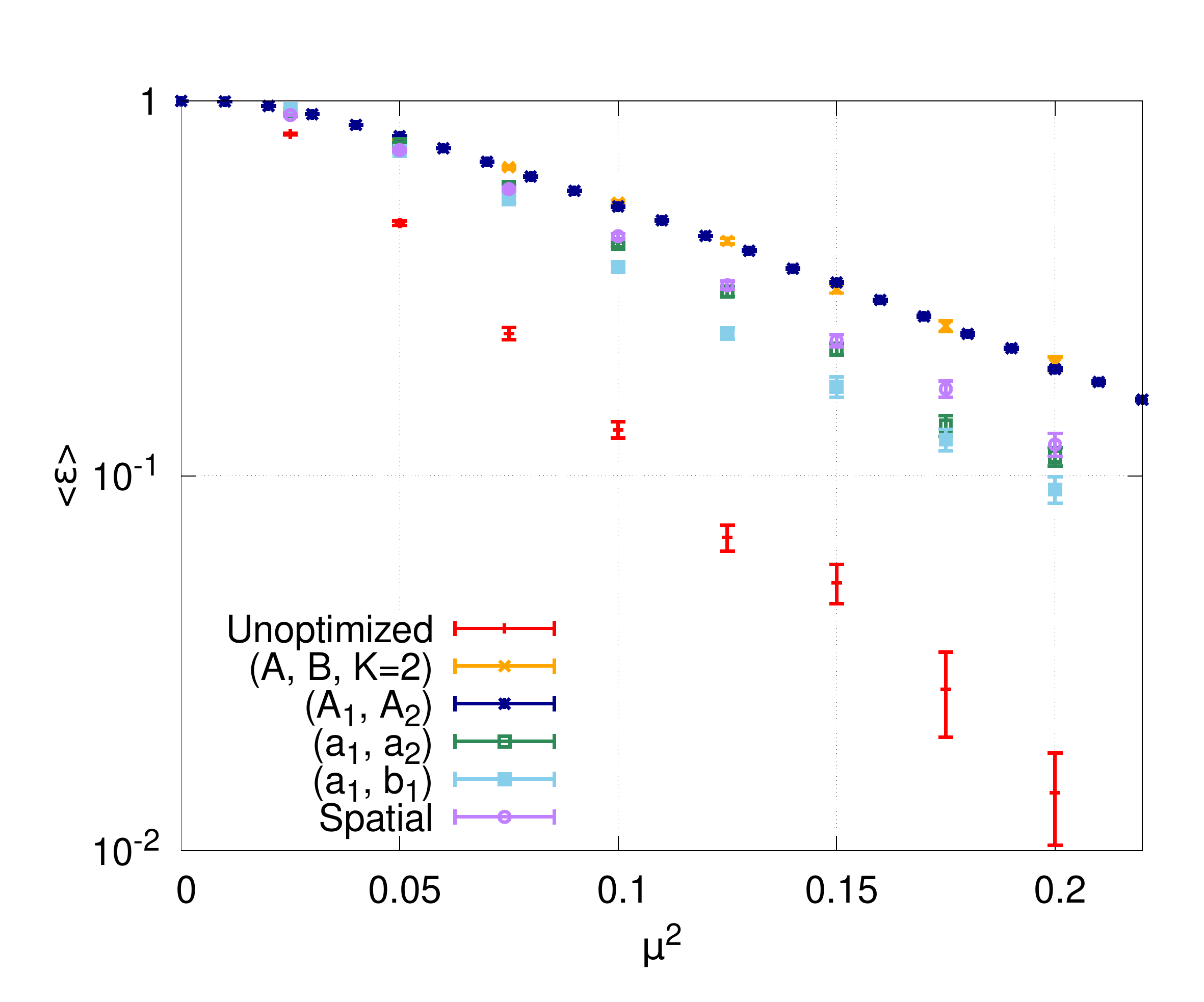}
    \caption{Severity of the sign problem on an $\Omega=8^3$ lattice with
      $\beta=0.4$. Unimproved results are shown in red. The optimized
      result achieved with the parametrizations given in
      Eq.~(\ref{fullparam}) with $K=2$ are shown in orange. The optimized
      results achieved with Eq.~(\ref{cutparam}) for $K=2$ are shown in purple. 
      The result from Eq.~(\ref{cutparam2}) with $a_1$ and $a_2$ non-zero are shown in green, while with $a_1$ and $b_1$ 
    non-zero are shown in light blue. Results with the fully translationally invariant $(A_1,A_2)$ parametrization are shown in blue.}
    \label{fig:comppar}
    \end{figure}

Since the chemical potential affects directly only the interaction
between temporal nearest neighbors, it is reasonable to consider
deformed manifolds that depend only on their difference. Given the
translation invariance of the system, it is also natural to expect
that the optimal deformation is also translationally invariant. For
more generality, we allow for a non-translationally invariant optimum
in the temporal direction, using the same coefficients for every
$\varphi_x$ with the same temporal coordinate $x_0$. We later check
our assumptions on spatial translation invariance and on neglecting
spatial neighbors. We then considered parametrizations of the
following general form,

 \begin{equation}
   \label{fullparam}
   \begin{aligned}
 \varphi_x(t_x,t_{x+\hat{0}}) &= t_x + \imath\Big\{
 A_{0,x_0} \\ &+\sum_{k=1}^K\Big[  A_{k,x_0}\,\cos(k(t_x-t_{x+\hat{0}}))
 \\ &+ B_{k,x_0}\, \sin(k(t_x-t_{x+\hat{0}})) \Big] \Big\},
   \end{aligned}
 \end{equation}
leading to the following effective action:
\begin{align}
S_{\mathrm{eff}} &= S - \sum_x\ln\frac{\partial\varphi_x}{\partial t_x} \nonumber \\
&\phantom{=}- \sum_{x_1,x_2}\ln\Bigg(1-(-1)^{N_0}\prod_{x_0}\frac{1-\partial\varphi_x/\partial t_x}{\partial\varphi_x/\partial t_x}\Bigg).
\end{align}
We used the choices $K=1$ and $2$ for the cut-off on the Fourier
series, which we will denote with $(A,B,K=1)$ and $(A,B,K=2)$, respectively. 
Performing optimization on a lattice of size $\Omega\equiv N_0N_1N_2=8^3$ for $\beta=0.4$ we find that the sign problem is substantially improved. In each iteration $10^6$ configurations were generated to compute the gradient. 
The initial values for the Fourier
coefficients were chosen to be zero for the smallest simulated $\mu^2$ value, and for each subsequent $\mu^2$ to be equal to the final values obtained in the previous optimization round.

Figure~\ref{fig:3Dsign} shows the average sign achieved with optimization for $K=1$ and $2$, along with the unoptimized results for comparison. There is significant improvement in the sign problem even when using only a first-order Fourier series. When going up to second order there is a marginal increase in $\langle\varepsilon\rangle_{\mathrm{SQ},\mu}$. The optimal values of the Fourier coefficients for $\mu^2=0.15$ and $K=2$ 
are listed in Tabs.~\ref{tab:3DK2coefs4n6} and \ref{tab:3DK2coefs8n10}
of Appendix~\ref{appx:tables}.

The coefficients of the constant and of the sine terms are 
two-three orders of magnitude smaller than the coefficients of the
cosine terms; they also fluctuate around zero as a function of $x_0$,
with a standard deviation larger than their average, while the
coefficients of the cosine terms have roughly the same value on every
time slice, with $A_{1,x_0}<0$ and $A_{2,x_0}>0$. 
This remains true at every simulated value of $\mu^2$.
This suggests that there is little gain in allowing for
$x_0$-dependent Fourier coefficients.
Furthermore, carrying out the optimization on $\Omega=8\times4^2,8\times6^2$ and $8\times10^2$ lattices we have found that the values of the optimal coefficients are close to those obtained on the $\Omega=8^3$ lattice.

\subsection{Ans\"atze with a triangular Jacobian matrix}

With the appearance of additional parameters the calculation of the
Jacobian generally becomes more involved.  Especially interesting from
the computational point of view are the parametrizations which keep
the Jacobian simple. For this reason, we considered ans\"atze with a
triangular Jacobian matrix (see also Ref.~\cite{Detmold:2021ulb}),
that have the benefit of having a very simple form of the effective
action :
\begin{equation}
S_{\mathrm{eff}} = S - \sum_x\ln\frac{\partial\varphi_x}{\partial t_x}.
\end{equation}
As we show below, this can be achieved at the price of losing
translational invariance.  
This is expected to reduce the amount of improvement that can be
gained by path deformation, 
especially at larger $\mu$~\cite{Bursa:2021org}. Nonetheless, there
may be a trade off with computational costs when the Jacobian gets
very complicated, in particular in systems where the dependence on
$\mu$ becomes milder at large $\mu$.

We considered 3 different ans\"atze of this type. First, we set
\begin{equation}
  \label{cutparam}
  \begin{aligned}
\varphi_x(&t_x,t_{x+\hat{0}},t_{x+\hat{1}},t_{x+\hat{2}}) =  \\
&t_x + \imath\sum_{k=1}^K\sum_{n=0}^2
\theta_n^{(1)}\,\Big[\bar{A}_{k,n}\,\cos(k(t_x-t_{x+\hat{n}}))  \\
&\phantom{t_x + \textstyle\imath\sum_{k=1}^K\,\,
\theta_n^{(1)}\,\Big[}
+ \bar{B}_{k,n}\,\sin(k(t_x-t_{x+\hat{n}}))\Big],
\end{aligned}
\end{equation}
where 
\begin{equation}
\theta_i^{(j)}=
\begin{cases}
1 &\quad \text{if } x_i<N_i-j, \\
0 &\quad \text{otherwise}.
\end{cases}
\end{equation}
Here, we constrained the $\varphi_x$ on the slices $x_i=N_i-1$ for
$i=0,1,2$ to stay independent of the real parts of their neighbors in
the $i$th direction, so as to make the Jacobian matrix triangular.
Except for these points we used global Fourier coefficients, i.e., the
same coefficients across all the other lattice sites. This leads to
the cancellation of the constant term since the action depends only on
the difference of the field variables.  The parametrization
Eq.~\eqref{cutparam} allows us to check the consequences of explicitly
breaking temporal translation invariance, and to assess whether the
optimal choice of contours is affected by the inclusion of
spatial neighbors in the parametrization.

We carried out optimization with this ansatz on lattices of sizes
$\Omega=8\times4^2, 8\times6^2, 8^3$ and $8\times10^2$ with
$\beta=0.4$. The optimal coefficients are shown in
Tab.~\ref{tab:3DK2coefspa} of Appendix~\ref{appx:tables} for
$\mu^2=0.15$ and $K=2$. We find that the optimal values of
$\bar{A}_{k,i}$ and $\bar{B}_{k,i}$ for $i=1,2$ and $\bar{B}_{k,0}$
are much smaller than those of $\bar{A}_{k,0}$, which in turn are in
agreement with those obtained for $A_{k,x_0}$ using the ansatz of
Eq.~(\ref{fullparam}). 

This suggests that the contribution of spatial neighbors can
be neglected in the parametrization. Once again, we observed that the
optimal coefficients are to a good approximation independent of the
spatial size of the system.

We have also investigated the effect of including the second nearest neighbor in the temporal direction in the parametrization of $\varphi_x$. Once again, constraints were imposed similarly to Eq.~(\ref{cutparam}) to make the Jacobian simple, and global coefficients were used, 
\begin{equation}
  \label{cutparam2}
  \begin{aligned}
\varphi_x(t_x,t_{x+\hat{0}},t_{x+2\hat{0}}) =&  \\
t_x + \imath\sum_{k=1}^K\Big[&
\theta_0^{(1)}\,a_k\,\cos(k(t_x-t_{x+\hat{0}}))
 \\
&+\theta_0^{(2)}\,b_k\,\cos(k(t_x-t_{x+2\hat{0}})) \Big].
\end{aligned}
\end{equation}
We omitted the sine terms since their coefficients remained close to
zero with the previously tested
parametrizations. 
Optimization was performed using Eq.~(\ref{cutparam2}) with two
setups: 
using $K=2$ and adjusting $a_1$ and $a_2$ with $b_k$s set to zero,
later referred to as $(a_1,a_2)$ optimization; or using $K=1$ and
adjusting both $a_1$ and $b_1$, that we call $(a_1,b_1)$ optimization.

In Fig.~\ref{fig:comppar} we compare the optimized results obtained with the parametrizations of Eqs.~\eqref{cutparam} and \eqref{cutparam2} 
with the unoptimized results, and with the optimized results previously obtained using Eq.~\eqref{fullparam}.
Comparing results obtained with Eq.~\eqref{cutparam} for $K=2$ and the
$(a_1,a_2)$ parametrization, it is clear that including spatial
neighbors leads only to a marginal improvement, at the cost of a much
more complicated Jacobian. It is also apparent that including the second-order
term $a_2$ gives a larger increase in the optimized
$\langle\varepsilon\rangle_{\mathrm{SQ},\mu}$ in the range of the
simulated $\mu$ values, than including the first-order term $b_1$.
Hence, going to second order in the Fourier series is more important than including second neighbors in the parametrization.

\begin{figure*}[t!]
    \centering
    \includegraphics[width=0.48\textwidth]{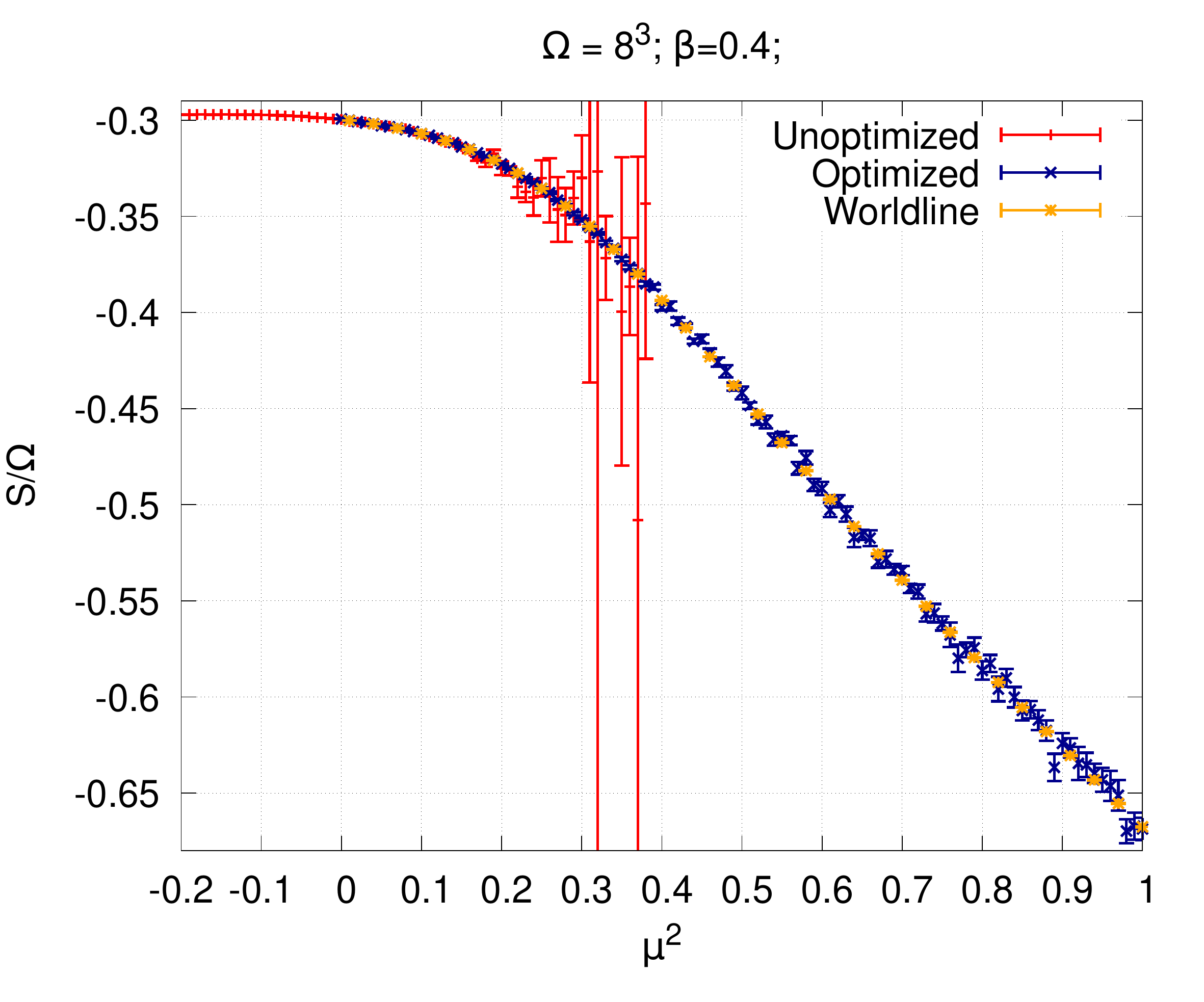}
    \includegraphics[width=0.48\textwidth]{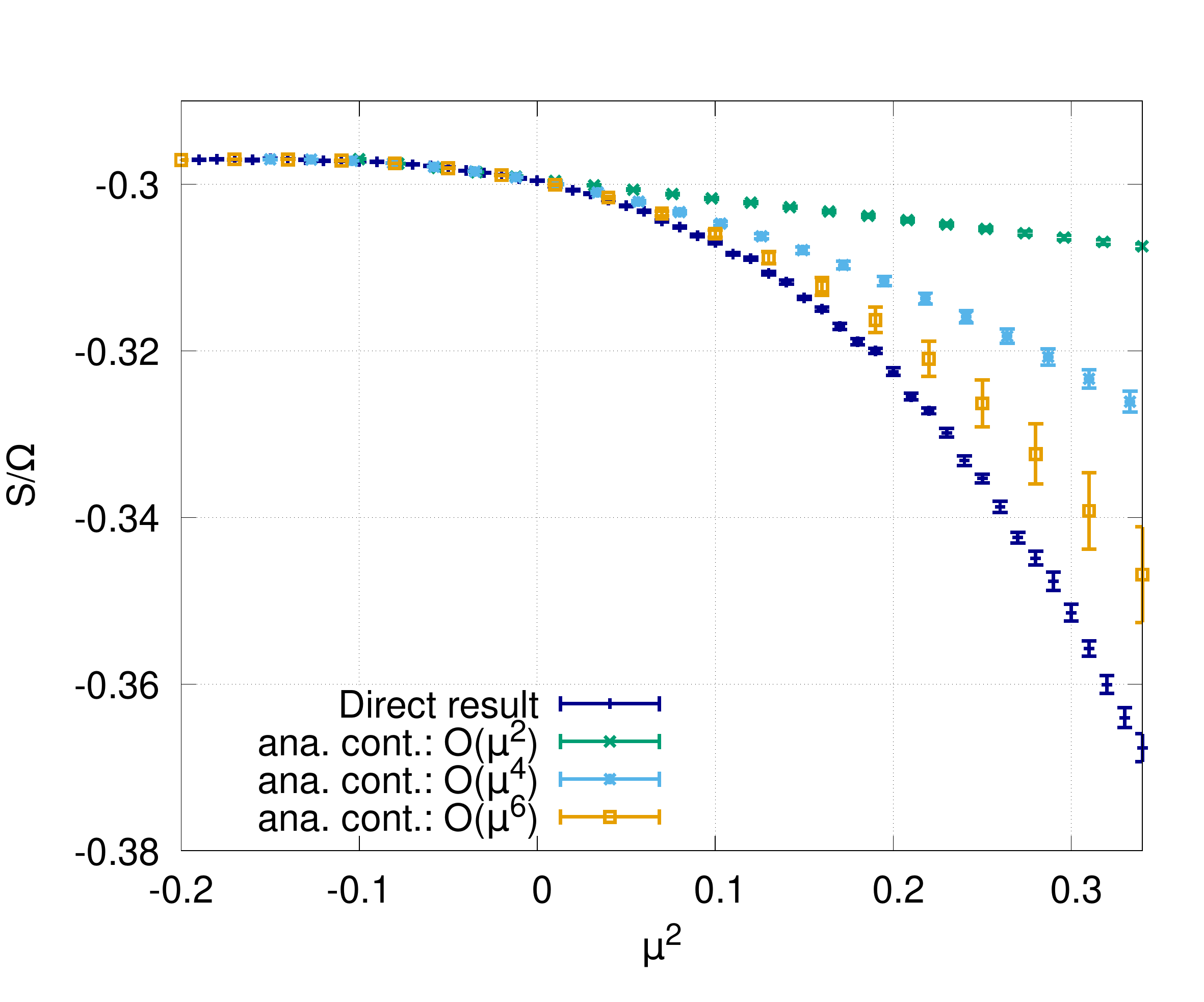}
    \caption{Left: Average action density on an $\Omega=8^3$ lattice with
      $\beta=0.4$. The unoptimized results, including those obtained with a local heat bath at $\mu^2\le 0$, are shown in red, the ($A_1,A_2)$ -optimized results in blue, and the worldline results in orange. Right: Average action density with the $(A_1,A_2)$-optimized contours, compared with 
      analytic continuation with polynomial ansätze up to orders $\mu^2$, $\mu^4$ and $\mu^6$ for small 
      chemical potentials.}
    \label{fig:action}
\end{figure*}

It is also clear that our parametrizations with a triangular Jacobian
matrix lead to a significant but not overwhelming loss in the improvement
of the sign problem compared to the ansatz of Eq.~(\ref{fullparam}). This
discrepancy can probably be attributed to the constraints introduced
in Eqs.~(\ref{cutparam}) and (\ref{cutparam2}). This is similar to
the conclusions of Ref.~\cite{Bursa:2021org} for Bose gases. 
In our model, parametrizations with the triangular Jacobian 
are not significantly cheaper to simulate the fully translationally invariant ansatz,
so there is no gain in breaking translational invariance at the boundaries of the lattice,
due to the sizeable difference in the improvement achieved.
In other models, there might be a less obvious trade-off.

With all of our two-parameter ans\"atze, scans of $\langle\varepsilon\rangle_{\mathrm{SQ},\mu}$ in the space of Fourier coefficients reveal a rather simple landscape. There is a single optimum located on a small plateau where the average sign changes slowly. 
As an example, scans of the average sign in the space of coefficients $a_1$ and $a_2$ are shown in Appendix~\ref{appx:scans}. 

\subsection{Fully translationally invariant ansatz}

Among our ans\"atze, the one that achieved the greatest improvement on the sign problem with the least amount 
of parameters was given by a translationally invariant version of Eq.~\eqref{fullparam}, with $K=2$, 
the constant and sine coefficients set to zero, and $A_{k,x_0} = A_k$
for all $x_0$. This will be denoted as the $(A_1,A_2)$ parametrization.

\section{Volume and chemical potential dependence of the sign problem on optimized manifolds}
\label{sect:results}

As a sanity check, we calculated the average action density as 
\begin{equation}
\frac{\langle S\rangle}{\Omega} = -\frac{\beta}{\Omega}\frac{\partial}{\partial\beta}\ln Z 
= \frac{ \langle\varepsilon(S^R + S^I \tan\Seff^I)\rangle_{\mathrm{SQ},\mu} }
{ \Omega\,\langle\varepsilon\rangle_{\mathrm{SQ},\mu} }.
\end{equation}
We first present the unoptimized and the $(A_1,A_2)$-optimized average action density in Fig.~\ref{fig:action}.
In both cases $10^8$ configurations were used. In the left panel of Fig.~\ref{fig:action} we also compare with results from the sign problem-free worldline formalism.  
We see good agreement between the different predictions. 
Predictions with the $(a_1,a_2)$-optimized scheme also agree in the 
range where the sign problem of the scheme is manageable.
In the right panel of Fig.~\ref{fig:action} we also compare with analytic continuation from $\mu^2<0$ with polynomial
ansätze of increasing order. The expansion converges rather slowly, even at small chemical potentials, way before the phase 
transition to the ordered phase.

\begin{figure*}[t!]
    \centering
    \includegraphics[width=0.48\textwidth]{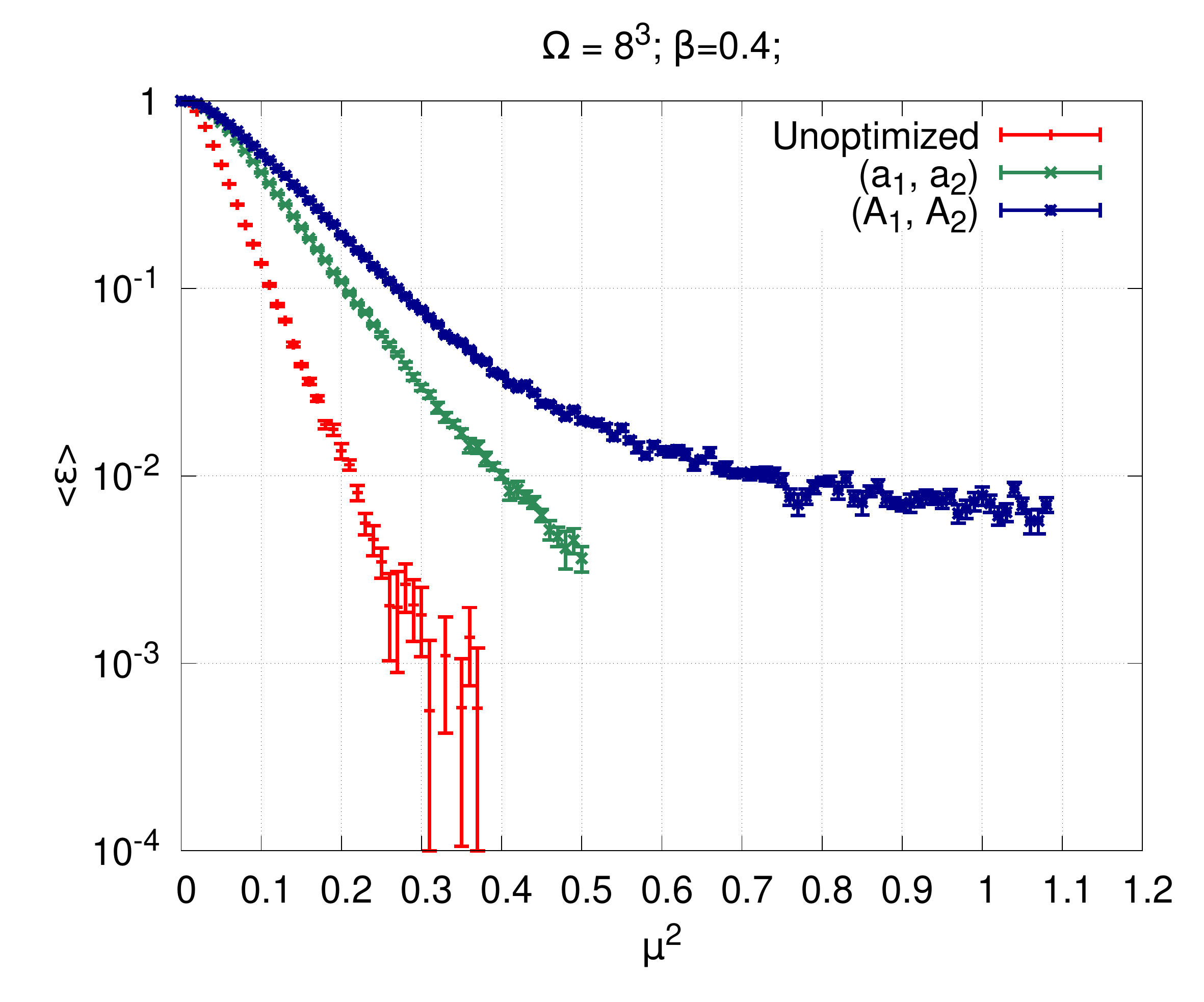}
    \includegraphics[width=0.48\textwidth]{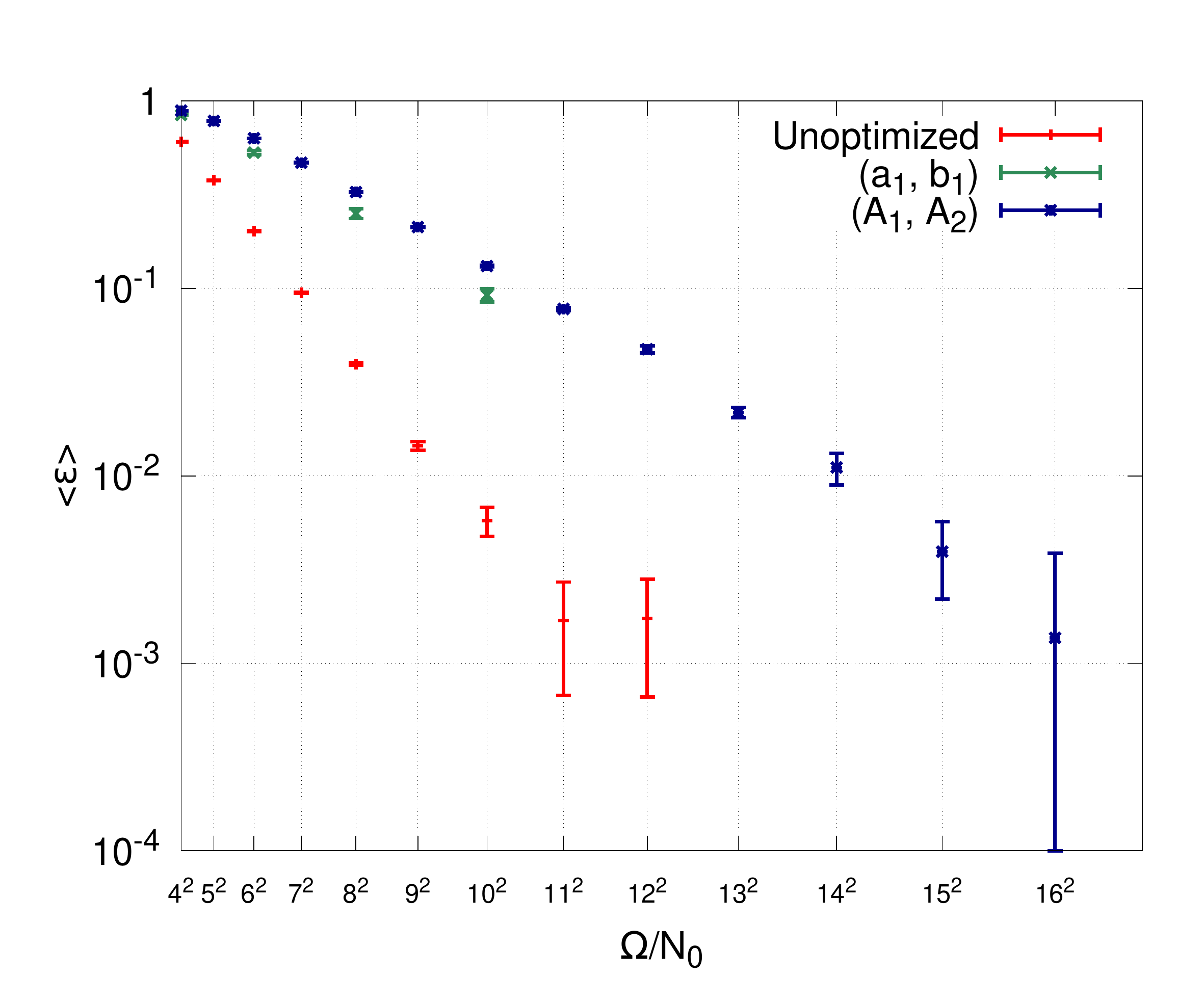}

    \caption{Left: Dependence on $\mu^2$ of the average sign for $\beta = 0.4$ and $\Omega = 8^3$ for the unoptimized calculations (red), and for the optimized calculations with $(a_1 , a_2 )$ (green) and $(A_1 , A_2 )$ (blue) parametrizations. Right: Volume dependence of the unoptimized (red) and $(a_1,a_2)$-optimized (green) and $(A_1,A_2)$-optimized (blue)    average signs. Results were obtained at $\beta=0.4$, $\mu^2=0.15$ with the temporal size fixed at $N_0=8$.}
    \label{fig:signs}
\end{figure*}

The significant improvement of the sign problem achieved by path
optimization is clearly visible in the left panel of Fig.~\ref{fig:signs}, where we
compare the average sign in the unoptimized case and in the optimized
case with parametrizations $(a_1,a_2)$ and $(A_1,A_2)$. The ratio of
this quantity between our best parametrization and the unoptimized
case is already of order $10^2$ at $\mu^2=0.3$, beyond which the
unoptimized approach fails. For the $(A_1,A_2)$ parametrization, the
path optimization method instead works well also deep in the ordered
phase at $\mu^2>\mu_c^2\approx 0.54$. An exponential fit
$\langle\varepsilon\rangle_{\mathrm{SQ},\mu}\sim e^{-C^{(\mu)}\mu^2}$
in the range $[0.1,0.25]$ yields
$C^{(\mu)}_{\mathrm{unopt}}\approx 24$,
$C^{(\mu)}_{(a_1,a_2)}\approx 13$, and
$C^{(\mu)}_{(A_1,A_2)}\approx 10$.  Notice that since the sign problem
should become mild at large $\mu$, we expect the average sign to reach
a minimum as a function of $\mu^2$ (at fixed volume), and then 
increase. This probably explains the flattening of the $(A_1,A_2)$ curve,
starting from around $\mu^2\approx 0.4$, which is likely related to
the transition to the ordered phase. An eyeball estimate leads to
expect several orders of magnitude of improvement in the central
region where the sign problem is at its strongest.

In the right panel of Fig.~\ref{fig:signs} we show the dependence of the average sign on
the volume for the unoptimized and for the $(a_1,a_2)$ and
$(A_1,A_2)$-optimized cases. A clear exponential decrease
$\langle\varepsilon\rangle_{\mathrm{SQ},\mu}\sim e^{-C^{(V)}V}$ is visible,
with $C^{(V)}_{\mathrm{unopt}}\approx 0.0073$,
$C^{(V)}_{(a_1,a_2)}\approx 0.0032$, and
$C^{(V)}_{(A_1,A_2)}\approx 0.0031$.  The improvement of the sign
problem by path optimization is exponential in the volume, reducing
the ``badness'' $C^{(V)}$ of the volume scaling by 50\%.

\section{Optimization with reweighting}
\label{sect:rew}

Generating new configurations at every optimization step can be computationally too expensive in more complicated models. A possible way to decrease the computational cost is to generate a set of configurations before starting the optimization procedure, and use the same set at every step to compute the gradient of the cost function through reweighting.
This would allow us to save time on the generation of configurations. 
When computing expectation values we need to reweight from the original weights, $r=|\cos\Seffo^I|\,e^{-\Seffo^R}$, to the new weights, $w=|\cos\Seffi^I|\,e^{-\Seffi^R}$, where $\Seffo$ 
is evaluated on a fixed set of configurations, e.g., those obtained
in the optimization at the previous $\mu^2$ value (or along the real axis for the smallest $\mu^2$), and $\Seffi$ is what we get with the updated values of the $p_i$ contour coefficients. The cost function is computed as
\begin{equation}
\frac{N_-}{N_+} = \frac{\langle\Theta(-\cos\Seffi^I)\,\frac{w}{r}\rangle_r}{\langle\Theta(\cos\Seffi^I)\,\frac{w}{r}\rangle_r},
\end{equation}
and its gradient with respect to the contour coefficients is
  \begin{equation}
    \label{eq:opt_rew}
    \begin{aligned}
\frac{\partial}{\partial p_i}\left(\frac{N_-}{N_+}\right) = \frac{N_-}{N_+}\Bigg[
&\frac{\langle\Theta(-\cos\Seffi^I)\,F_i\frac{w}{r}\rangle_r}{\langle\Theta(-\cos\Seffi^I)\,\frac{w}{r}\rangle_r}
\\ &-\frac{\langle\Theta(\cos\Seffi^I)\,F_i\frac{w}{r}\rangle_r}{\langle\Theta(\cos\Seffi^I)\,\frac{w}{r}\rangle_r}
\Bigg],
    \end{aligned}
\end{equation}
where $\langle\dots\rangle_r$ denotes averaging with respect to the original weights $r$. One downside of this method is that it might introduce an overlap problem which we need to monitor throughout the optimization procedure. 
This can be done by looking at the numerical results for the denominators in
Eq.~\eqref{eq:opt_rew}.
When these fall below a prescribed tolerance level and the overlap
problem becomes too severe, we generate a new set of
configurations before proceeding with the updates. Note that this overlap problem 
in the optimization by no means can bias the final results, since even on 
unoptimal contours the integral is guaranteed to be the same by the multi-dimensional 
Cauchy theorem. On the other hand, it could lead to a loss in the improvement 
on the sign problem: that is why the second step of the procedure - generating new 
configurations when the overlap problem becomes too severe - is useful.

\begin{figure}[htbp]
\centering
\includegraphics[width=0.48\textwidth]{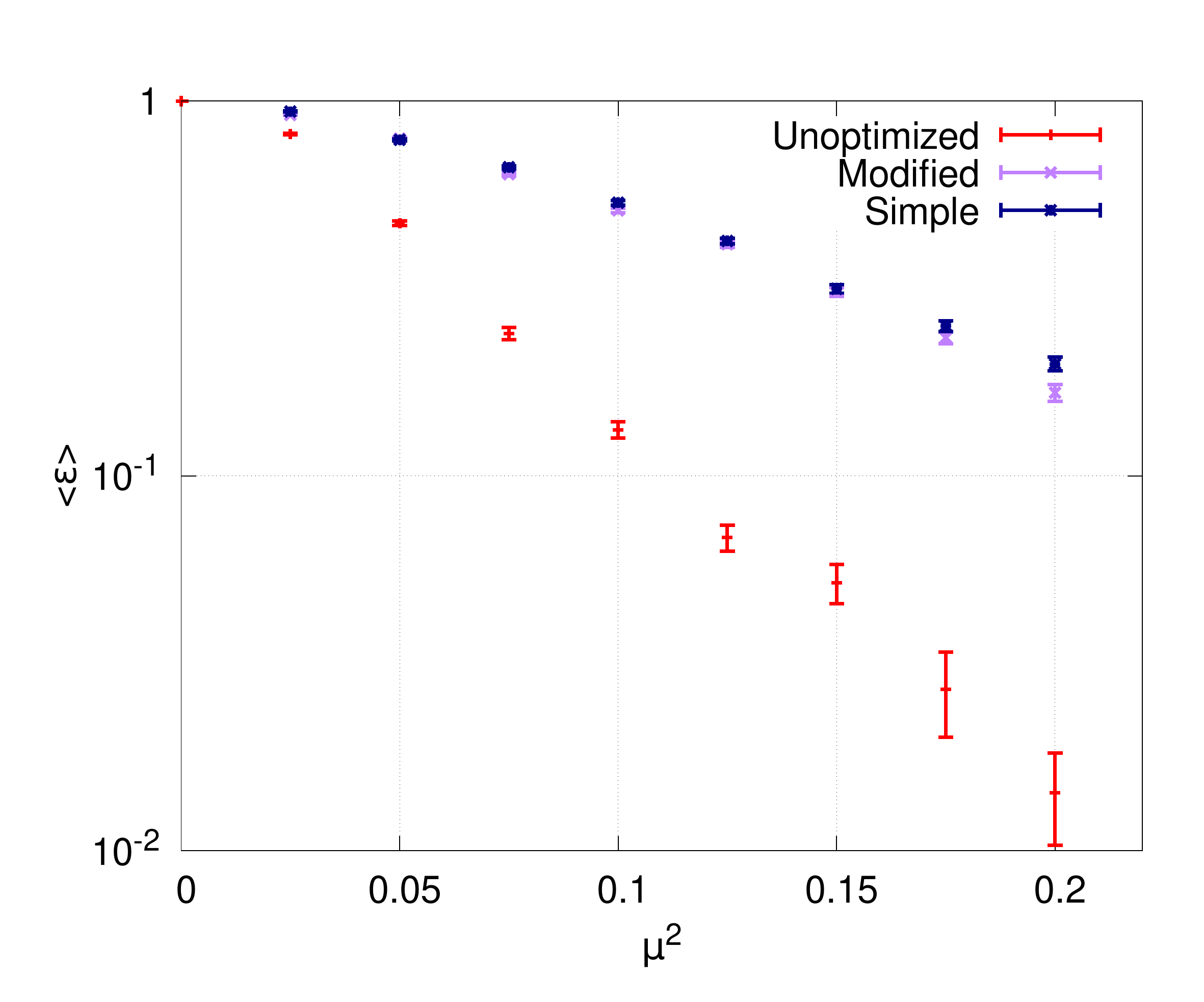}
\caption{Comparison of the average sign achieved with the simple and the modified optimization method at $\beta=0.4$ and $\Omega=8^3$ with the 
ansatz of Eq.~\eqref{fullparam}.}
\label{fig:rewop1}
\end{figure}

As it can be seen in  Fig.~\ref{fig:rewop1}, this modified optimization method yielded similarly good results when compared to the procedure used
previously, which required the generation of new configurations at every step of the iteration. Here we used the parametrization of Eq.~(\ref{fullparam}). We observed a good agreement between the values of the coefficients obtained with the simple and the modified optimization. The optimal values for $\beta=0.4$, $\Omega=8^3$ and $\mu^2=0.15$ are shown in Tab.~\ref{tab:rewcoef0p15} of Appendix~\ref{appx:tables}. 

\section{Discussion}
\label{sect:disc}

In this paper we studied the path optimization method for reducing the severity of the sign problem in the $2+1$ dimensional XY model with nonzero chemical potential. We used simple parametrizations for the complexified field variables. 
We have shown that the optimized manifold 
exhibits an explicit temporal translational invariance. 
Exploiting this property allows us to use significantly fewer optimization parameters.

Furthermore, we have found that the optimal choice of contours appears to be independent of the spatial size of the lattice. Such a feature can be utilized to make the optimization procedure computationally less expensive as it would be sufficient to find the optimal contours for a small lattice. Then simulations can be carried out for larger lattices using the same contours.

We have shown numerical evidence that the reduction of the sign problem is exponential both in the 
chemical potential and the volume - i.e. it considerably reduces the exponents characterizing its severity.
This was achieved without changing the number of parameters with the volume, keeping the number of optimizable parameters 
at a small fixed value.

On the optimized integration manifolds, it was possible to simulate also on the other side of the transition to the ordered phase.

We have also demonstrated that it is sufficient to generate
configurations only at the start of the optimization procedure. Then,
the same set of configurations is used to compute the gradient of the
cost function at each step. As the contours are updated, it is necessary to
reweight from the generated distribution to the one corresponding to
the new contours. When the overlap between the two distributions
significantly decreases, it is preferable to generate new configurations, in order not to lose optimizing power. With
this approach the computation time of the optimization can be
significantly reduced as compared to a method where new configurations
are generated at every iteration,
with no significant loss in the reduction of the severity of the sign problem.

\section*{Acknowledgements}
This work was partly
supported by the NKFIH grant KKP-126769. K.K. was supported by the
\'UNKP-21-3-II-ELTE-625 New National Excellence Program of the
Ministry of Human Capacities of Hungary. A.P. is supported by the J. Bolyai Research
Scholarship of the Hungarian Academy of Sciences and by the \'UNKP-21-5 New
National Excellence Program of the Ministry for Innovation and Technology from the 
source of the National Research, Development and Innovation Fund.

\clearpage
\appendix
\section{Optimal coefficient values}
\label{appx:tables}

Here we present the optimal Fourier coefficients obtained at $\beta=0.4$ and $\mu^2=0.15$ on $\Omega=8\times4^2,8\times6^2,8^3$ and $8\times10^2$ lattices for $K=2.$ The optimal values for the parametrization described by Eq.~(\ref{fullparam}) are given in Tabs.~\ref{tab:3DK2coefs4n6} and \ref{tab:3DK2coefs8n10} while Tab.~\ref{tab:3DK2coefspa} shows the ones for Eq.~(\ref{cutparam}). We compare the coefficients obtained with the reweighting technique introduced in Sec.~\ref{sect:rew} to those that were found using simple optimization in Tab.~\ref{tab:rewcoef0p15}. 

\begin{widetext}

\begin{table*}[!htbp]
\centering
\begin{tabular}{|c|c|c||c|c|c|} \hline
term & value, $8\times4^2$ & value, $8\times6^2$ & term & value, $8\times4^2$ & value, $8\times6^2$ \\ \hline
$A_{0,0}$ &  $ 3.68\times10^{-3}$ &  $-2.96\times10^{-3}$ & $A_{0,4}$ &  $ 6.15\times10^{-4}$ &  $-1.18\times10^{-3}$ \\ 
$\boldsymbol{A_{1,0}}$ &  $\boldsymbol{-1.08\times10^{-1}}$ &  $\boldsymbol{-1.19\times10^{-1}}$ & $\boldsymbol{A_{1,4}}$ &  $\boldsymbol{-1.12\times10^{-1}}$ &  $\boldsymbol{-1.15\times10^{-1}}$ \\ 
$B_{1,0}$ &  $-6.57\times10^{-4}$ &  $-9.40\times10^{-4}$ & $B_{1,4}$ &  $-9.48\times10^{-4}$ &  $-1.90\times10^{-3}$ \\ 
$\boldsymbol{A_{2,0}}$ &  $ \boldsymbol{1.34\times10^{-2}}$ &  $ \boldsymbol{1.29\times10^{-2}}$ & $\boldsymbol{A_{2,4}}$ &  $ \boldsymbol{1.29\times10^{-2}}$ &  $ \boldsymbol{1.13\times10^{-2}}$ \\ 
$B_{2,0}$ &  $ 1.77\times10^{-3}$ &  $-1.79\times10^{-3}$ & $B_{2,4}$ &  $ 6.92\times10^{-4}$ &  $-2.93\times10^{-4}$ \\ \hline 
$A_{0,1}$ &  $ 4.35\times10^{-3}$ &  $-4.15\times10^{-3}$ & $A_{0,5}$ &  $-5.45\times10^{-3}$ &  $ 1.69\times10^{-3}$ \\ 
$\boldsymbol{A_{1,1}}$ &  $\boldsymbol{-1.15\times10^{-1}}$ &  $\boldsymbol{-1.13\times10^{-1}}$ & $\boldsymbol{A_{1,5}}$ &  $\boldsymbol{-1.06\times10^{-1}}$ &  $\boldsymbol{-1.14\times10^{-1}}$ \\ 
$B_{1,1}$ &  $ 8.11\times10^{-4}$ &  $-1.24\times10^{-3}$ & $B_{1,5}$ &  $-5.92\times10^{-4}$ &  $ 1.86\times10^{-3}$ \\ 
$\boldsymbol{A_{2,1}}$ &  $ \boldsymbol{1.37\times10^{-2}}$ &  $ \boldsymbol{1.36\times10^{-2}}$ & $\boldsymbol{A_{2,5}}$ &  $ \boldsymbol{1.19\times10^{-2}}$ &  $ \boldsymbol{1.30\times10^{-2}}$ \\ 
$B_{2,1}$ &  $-1.26\times10^{-3}$ &  $-5.67\times10^{-4}$ & $B_{2,5}$ &  $-5.27\times10^{-4}$ &  $-1.33\times10^{-4}$ \\ \hline 
$A_{0,2}$ &  $-3.66\times10^{-3}$ &  $ 3.18\times10^{-3}$ & $A_{0,6}$ &  $ 2.53\times10^{-3}$ &  $ 2.42\times10^{-3}$ \\ 
$\boldsymbol{A_{1,2}}$ &  $\boldsymbol{-1.11\times10^{-1}}$ &  $\boldsymbol{-1.19\times10^{-1}}$ & $\boldsymbol{A_{1,6}}$ &  $\boldsymbol{-1.20\times10^{-1}}$ &  $\boldsymbol{-1.14\times10^{-1}}$ \\ 
$B_{1,2}$ &  $-7.33\times10^{-4}$ &  $-4.09\times10^{-4}$ & $B_{1,6}$ &  $ 8.31\times10^{-4}$ &  $-2.00\times10^{-3}$ \\ 
$\boldsymbol{A_{2,2}}$ &  $ \boldsymbol{1.26\times10^{-2}}$ &  $ \boldsymbol{1.28\times10^{-2}}$ & $\boldsymbol{A_{2,6}}$ &  $ \boldsymbol{1.44\times10^{-2}}$ &  $ \boldsymbol{1.23\times10^{-2}}$ \\ 
$B_{2,2}$ &  $ 2.64\times10^{-4}$ &  $-7.20\times10^{-4}$ & $B_{2,6}$ &  $-1.36\times10^{-3}$ &  $ 4.47\times10^{-5}$ \\ \hline 
$A_{0,3}$ &  $ 3.88\times10^{-3}$ &  $ 2.35\times10^{-3}$ & $A_{0,7}$ &  $-4.80\times10^{-3}$ &  $ 4.61\times10^{-3}$ \\ 
$\boldsymbol{A_{1,3}}$ &  $\boldsymbol{-1.16\times10^{-1}}$ &  $\boldsymbol{-1.18\times10^{-1}}$ & $\boldsymbol{A_{1,7}}$ &  $\boldsymbol{-1.09\times10^{-1}}$ &  $\boldsymbol{-1.20\times10^{-1}}$ \\ 
$B_{1,3}$ &  $-2.13\times10^{-3}$ &  $-2.14\times10^{-3}$ & $B_{1,7}$ &  $-2.58\times10^{-3}$ &  $ 1.24\times10^{-5}$ \\ 
$\boldsymbol{A_{2,3}}$ &  $ \boldsymbol{1.22\times10^{-2}}$ &  $ \boldsymbol{1.40\times10^{-2}}$ & $\boldsymbol{A_{2,7}}$ &  $ \boldsymbol{1.35\times10^{-2}}$ &  $ \boldsymbol{1.39\times10^{-2}}$ \\ 
$B_{2,3}$ &  $ 2.10\times10^{-4}$ &  $ 8.39\times10^{-4}$ & $B_{2,7}$ &  $-1.68\times10^{-3}$ &  $-1.04\times10^{-3}$ \\ \hline 
\end{tabular}
\caption{Optimal value of the Fourier coefficients obtained on $\Omega=8\times4^2$ and $8\times6^2$ lattices for $\beta=0.4$, $\mu^2=0.15$, $K=2$ with parametrization given by Eq.~(\ref{fullparam}). The most relevant values are written in bold while the others are negligible.}
\label{tab:3DK2coefs4n6}
\end{table*}

\begin{table*}[!htbp]
\centering
\begin{tabular}{|c|c|c||c|c|c|} \hline
term & value, $8^3$ & value, $8\times10^2$ & term & value, $8^3$ & value, $8\times10^2$ \\ \hline
$A_{0,0}$ &  $-1.47\times10^{-4}$ &  $ 3.77\times10^{-3}$ & $A_{0,4}$ &  $ 4.15\times10^{-4}$ &  $-2.28\times10^{-3}$ \\ 
$\boldsymbol{A_{1,0}}$ &  $\boldsymbol{-1.23\times10^{-1}}$ &  $\boldsymbol{-1.21\times10^{-1}}$ & $\boldsymbol{A_{1,4}}$ &  $\boldsymbol{-1.19\times10^{-1}}$ &  $\boldsymbol{-1.24\times10^{-1}}$ \\ 
$B_{1,0}$ &  $-3.05\times10^{-4}$ &  $ 1.10\times10^{-3}$ & $B_{1,4}$ &  $ 1.70\times10^{-3}$ &  $ 2.63\times10^{-3}$ \\ 
$\boldsymbol{A_{2,0}}$ &  $ \boldsymbol{1.54\times10^{-2}}$ &  $ \boldsymbol{1.37\times10^{-2}}$ & $\boldsymbol{A_{2,4}}$ &  $ \boldsymbol{1.38\times10^{-2}}$ &  $ \boldsymbol{1.29\times10^{-2}}$ \\ 
$B_{2,0}$ &  $ 5.83\times10^{-4}$ &  $-1.43\times10^{-4}$ & $B_{2,4}$ &  $-7.15\times10^{-5}$ &  $ 2.83\times10^{-4}$ \\ \hline 
$A_{0,1}$ &  $-1.33\times10^{-3}$ &  $-3.80\times10^{-3}$ & $A_{0,5}$ &  $-2.43\times10^{-4}$ &  $-2.34\times10^{-3}$ \\ 
$\boldsymbol{A_{1,1}}$ &  $\boldsymbol{-1.18\times10^{-1}}$ &  $\boldsymbol{-1.21\times10^{-1}}$ & $\boldsymbol{A_{1,5}}$ &  $\boldsymbol{-1.20\times10^{-1}}$ &  $\boldsymbol{-1.23\times10^{-1}}$ \\ 
$B_{1,1}$ &  $ 1.05\times10^{-3}$ &  $ 2.84\times10^{-3}$ & $B_{1,5}$ &  $-1.58\times10^{-3}$ &  $ 6.28\times10^{-4}$ \\ 
$\boldsymbol{A_{2,1}}$ &  $ \boldsymbol{1.66\times10^{-2}}$ &  $ \boldsymbol{1.58\times10^{-2}}$ & $\boldsymbol{A_{2,5}}$ &  $ \boldsymbol{1.21\times10^{-2}}$ &  $ \boldsymbol{1.13\times10^{-2}}$ \\ 
$B_{2,1}$ &  $ 1.56\times10^{-3}$ &  $ 1.78\times10^{-3}$ & $B_{2,5}$ &  $-1.64\times10^{-3}$ &  $-1.46\times10^{-5}$ \\ \hline 
$A_{0,2}$ &  $ 9.76\times10^{-3}$ &  $ 8.90\times10^{-3}$ & $A_{0,6}$ &  $-2.74\times10^{-3}$ &  $-4.49\times10^{-3}$ \\ 
$\boldsymbol{A_{1,2}}$ &  $\boldsymbol{-1.20\times10^{-1}}$ &  $\boldsymbol{-1.21\times10^{-1}}$ & $\boldsymbol{A_{1,6}}$ &  $\boldsymbol{-1.19\times10^{-1}}$ &  $\boldsymbol{-1.22\times10^{-1}}$ \\ 
$B_{1,2}$ &  $ 2.15\times10^{-3}$ &  $ 1.31\times10^{-3}$ & $B_{1,6}$ &  $ 9.43\times10^{-5}$ &  $-4.86\times10^{-4}$ \\ 
$\boldsymbol{A_{2,2}}$ &  $ \boldsymbol{1.47\times10^{-2}}$ &  $ \boldsymbol{1.61\times10^{-2}}$ & $\boldsymbol{A_{2,6}}$ &  $ \boldsymbol{1.69\times10^{-2}}$ &  $ \boldsymbol{1.42\times10^{-2}}$ \\ 
$B_{2,2}$ &  $ 2.10\times10^{-3}$ &  $-7.80\times10^{-4}$ & $B_{2,6}$ &  $ 6.84\times10^{-4}$ &  $-6.65\times10^{-4}$ \\ \hline 
$A_{0,3}$ &  $-4.54\times10^{-3}$ &  $-2.31\times10^{-3}$ & $A_{0,7}$ &  $ 5.67\times10^{-3}$ &  $ 3.58\times10^{-3}$ \\ 
$\boldsymbol{A_{1,3}}$ &  $\boldsymbol{-1.21\times10^{-1}}$ &  $\boldsymbol{-1.20\times10^{-1}}$ & $\boldsymbol{A_{1,7}}$ &  $\boldsymbol{-1.22\times10^{-1}}$ &  $\boldsymbol{-1.23\times10^{-1}}$ \\ 
$B_{1,3}$ &  $-2.61\times10^{-3}$ &  $-2.21\times10^{-3}$ & $B_{1,7}$ &  $ 1.49\times10^{-3}$ &  $ 7.52\times10^{-4}$ \\ 
$\boldsymbol{A_{2,3}}$ &  $ \boldsymbol{1.35\times10^{-2}}$ &  $ \boldsymbol{1.62\times10^{-2}}$ & $\boldsymbol{A_{2,7}}$ &  $ \boldsymbol{1.27\times10^{-2}}$ &  $ \boldsymbol{1.29\times10^{-2}}$ \\ 
$B_{2,3}$ &  $-3.84\times10^{-4}$ &  $-2.73\times10^{-4}$ & $B_{2,7}$ &  $ 6.48\times10^{-4}$ &  $ 4.98\times10^{-5}$ \\ \hline 
\end{tabular}
\caption{Optimal value of the Fourier coefficients obtained on an $\Omega=8^3$ lattice for $\beta=0.4$, $\mu^2=0.15$, $K=2$ with parametrization given by Eq.~(\ref{fullparam}). The most relevant values are written in bold while the others are negligible.}
\label{tab:3DK2coefs8n10}
\end{table*}

\begin{table*}[!htbp]
\centering
\begin{tabular}{|c|c|c|c|c|} \hline
term & value, $8\times4^2$ & value, $8\times6^2$ & value, $8^3$ & value, $8\times10^2$ \\ \hline
$\boldsymbol{\bar{A}_{1,0}}$ & $\boldsymbol{-1.15\times10^{-1}}$ & $\boldsymbol{-1.15\times10^{-1}}$ & $\boldsymbol{-1.18\times10^{-1}}$ & $\boldsymbol{-1.20\times10^{-1}}$ \\
$\bar{B}_{1,0}$ & $ 2.44\times10^{-3}$ & $-3.63\times10^{-4}$ & $-1.85\times10^{-3}$ & $-2.40\times10^{-3}$ \\
$\boldsymbol{\bar{A}_{2,0}}$ & $ \boldsymbol{1.35\times10^{-2}}$ & $ \boldsymbol{1.48\times10^{-2}}$ & $ \boldsymbol{1.41\times10^{-2}}$ & $ \boldsymbol{1.39\times10^{-2}}$ \\
$\bar{B}_{2,0}$ & $ 8.16\times10^{-5}$ & $-1.26\times10^{-3}$ & $-1.50\times10^{-3}$ & $ 1.57\times10^{-3}$ \\ \hline
$\bar{A}_{1,1}$ & $ 5.44\times10^{-3}$ & $ 6.13\times10^{-3}$ & $ 3.68\times10^{-3}$ & $ 3.60\times10^{-3}$ \\
$\bar{B}_{1,1}$ & $ 2.57\times10^{-3}$ & $-5.47\times10^{-4}$ & $-2.61\times10^{-4}$ & $ 2.42\times10^{-3}$ \\
$\bar{A}_{2,1}$ & $-5.32\times10^{-4}$ & $-1.92\times10^{-3}$ & $-5.14\times10^{-5}$ & $ 1.42\times10^{-3}$ \\
$\bar{B}_{2,1}$ & $ 5.13\times10^{-4}$ & $ 6.43\times10^{-5}$ & $ 1.06\times10^{-3}$ & $ 1.56\times10^{-4}$ \\ \hline
$\bar{A}_{1,2}$ & $ 6.04\times10^{-3}$ & $ 5.62\times10^{-3}$ & $ 5.16\times10^{-3}$ & $ 4.43\times10^{-3}$ \\
$\bar{B}_{1,2}$ & $ 2.22\times10^{-3}$ & $ 6.81\times10^{-4}$ & $-3.57\times10^{-4}$ & $-1.23\times10^{-3}$ \\
$\bar{A}_{2,2}$ & $ 2.48\times10^{-4}$ & $-1.13\times10^{-3}$ & $-5.08\times10^{-4}$ & $-1.40\times10^{-3}$ \\
$\bar{B}_{2,2}$ & $-3.67\times10^{-4}$ & $-7.64\times10^{-4}$ & $ 3.33\times10^{-4}$ & $ 1.15\times10^{-3}$ \\ \hline
\end{tabular}
\caption{Optimal values of the Fourier coefficients obtained on $\Omega=8\times4^2, 8\times6^2, 8^3$ and $8\times10^2$ lattices for $\beta=0.4$, $\mu^2=0.15$ and $K=2$ with parametrization given by Eq.~(\ref{cutparam}). The most relevant values are written in bold while the others are negligible.}
\label{tab:3DK2coefspa}
\end{table*}

\begin{table*}[!htbp]
\centering
\begin{tabular}{|c|c|c||c|c|c|} \hline
term & modified & simple & term & modified & simple \\ \hline
$A_{0,0}$ &  $-1.48\times10^{-4}$ &  $-5.66\times10^{-3}$ & $A_{0,4}$ &  $ 4.15\times10^{-4}$ &  $-3.06\times10^{-2}$ \\ 
$\boldsymbol{A_{1,0}}$ &  $\boldsymbol{-1.23\times10^{-1}}$ &  $\boldsymbol{-1.11\times10^{-1}}$ & $\boldsymbol{A_{1,4}}$ &  $\boldsymbol{-1.19\times10^{-1}}$ &  $\boldsymbol{-1.19\times10^{-1}}$ \\ 
$B_{1,0}$ &  $-3.06\times10^{-4}$ &  $ 1.66\times10^{-3}$ & $B_{1,4}$ &  $ 1.71\times10^{-3}$ &  $ 3.67\times10^{-3}$ \\ 
$\boldsymbol{A_{2,0}}$ &  $ \boldsymbol{1.55\times10^{-2}}$ &  $ \boldsymbol{1.03\times10^{-2}}$ & $\boldsymbol{A_{2,4}}$ &  $ \boldsymbol{1.38\times10^{-2}}$ &  $ \boldsymbol{1.03\times10^{-2}}$ \\ 
$B_{2,0}$ &  $ 5.83\times10^{-4}$ &  $ 2.40\times10^{-3}$ & $B_{2,4}$ &  $-7.15\times10^{-5}$ &  $-4.33\times10^{-3}$ \\ \hline 
$A_{0,1}$ &  $-1.33\times10^{-3}$ &  $-5.65\times10^{-3}$ & $A_{0,5}$ &  $-2.43\times10^{-4}$ &  $ 4.74\times10^{-3}$ \\ 
$\boldsymbol{A_{1,1}}$ &  $\boldsymbol{-1.19\times10^{-1}}$ &  $\boldsymbol{-1.09\times10^{-1}}$ & $\boldsymbol{A_{1,5}}$ &  $\boldsymbol{-1.20\times10^{-1}}$ &  $\boldsymbol{-1.21\times10^{-1}}$ \\ 
$B_{1,1}$ &  $ 1.05\times10^{-3}$ &  $ 2.51\times10^{-3}$ & $B_{1,5}$ &  $-1.58\times10^{-3}$ &  $-6.40\times10^{-3}$ \\ 
$\boldsymbol{A_{2,1}}$ &  $ \boldsymbol{1.66\times10^{-2}}$ &  $ \boldsymbol{1.44\times10^{-2}}$ & $\boldsymbol{A_{2,5}}$ &  $ \boldsymbol{1.22\times10^{-2}}$ &  $ \boldsymbol{1.63\times10^{-2}}$ \\ 
$B_{2,1}$ &  $ 1.57\times10^{-3}$ &  $-5.24\times10^{-3}$ & $B_{2,5}$ &  $-1.65\times10^{-3}$ &  $ 3.31\times10^{-4}$ \\ \hline 
$A_{0,2}$ &  $ 9.76\times10^{-3}$ &  $-3.26\times10^{-3}$ & $A_{0,6}$ &  $-2.75\times10^{-3}$ &  $-1.30\times10^{-3}$ \\ 
$\boldsymbol{A_{1,2}}$ &  $\boldsymbol{-1.21\times10^{-1}}$ &  $\boldsymbol{-1.27\times10^{-1}}$ & $\boldsymbol{A_{1,6}}$ &  $\boldsymbol{-1.20\times10^{-1}}$ &  $\boldsymbol{-1.04\times10^{-1}}$ \\ 
$B_{1,2}$ &  $ 2.15\times10^{-3}$ &  $-1.21\times10^{-2}$ & $B_{1,6}$ &  $ 9.44\times10^{-5}$ &  $ 1.26\times10^{-3}$ \\ 
$\boldsymbol{A_{2,2}}$ &  $ \boldsymbol{1.47\times10^{-2}}$ &  $ \boldsymbol{1.81\times10^{-2}}$ & $\boldsymbol{A_{2,6}}$ &  $ \boldsymbol{1.70\times10^{-2}}$ &  $ \boldsymbol{6.71\times10^{-3}}$ \\ 
$B_{2,2}$ &  $ 2.11\times10^{-3}$ &  $ 8.61\times10^{-3}$ & $B_{2,6}$ &  $ 6.84\times10^{-4}$ &  $ 1.36\times10^{-4}$ \\ \hline 
$A_{0,3}$ &  $-4.55\times10^{-3}$ &  $-2.27\times10^{-2}$ & $A_{0,7}$ &  $ 5.68\times10^{-3}$ &  $ 1.61\times10^{-3}$ \\ 
$\boldsymbol{A_{1,3}}$ &  $\boldsymbol{-1.22\times10^{-1}}$ &  $\boldsymbol{-1.22\times10^{-1}}$ & $\boldsymbol{A_{1,7}}$ &  $\boldsymbol{-1.22\times10^{-1}}$ &  $\boldsymbol{-1.18\times10^{-1}}$ \\ 
$B_{1,3}$ &  $-2.62\times10^{-3}$ &  $ 9.84\times10^{-3}$ & $B_{1,7}$ &  $ 1.49\times10^{-3}$ &  $-1.48\times10^{-3}$ \\ 
$\boldsymbol{A_{2,3}}$ &  $ \boldsymbol{1.36\times10^{-2}}$ &  $ \boldsymbol{1.28\times10^{-2}}$ & $\boldsymbol{A_{2,7}}$ &  $ \boldsymbol{1.28\times10^{-2}}$ &  $ \boldsymbol{1.36\times10^{-2}}$ \\ 
$B_{2,3}$ &  $-3.84\times10^{-4}$ &  $ 7.24\times10^{-4}$ & $B_{2,7}$ &  $ 6.49\times10^{-4}$ &  $ 6.42\times10^{-3}$ \\ \hline 
\end{tabular}
\caption{Comparison of the values of Fourier coefficients obtained with the simple and the reweighted optimization method on an $\Omega=8^3$ lattice for $\beta=0.4$, $\mu^2=0.15$ and $K=2$ with parametrization given by Eq.~(\ref{fullparam}).}
\label{tab:rewcoef0p15}
\end{table*}
\end{widetext}

\clearpage
\section{Scans in coefficient space}
\label{appx:scans}

\begin{widetext}
\begin{figure*}[htbp]
\centering
\includegraphics[width=0.45\textwidth]{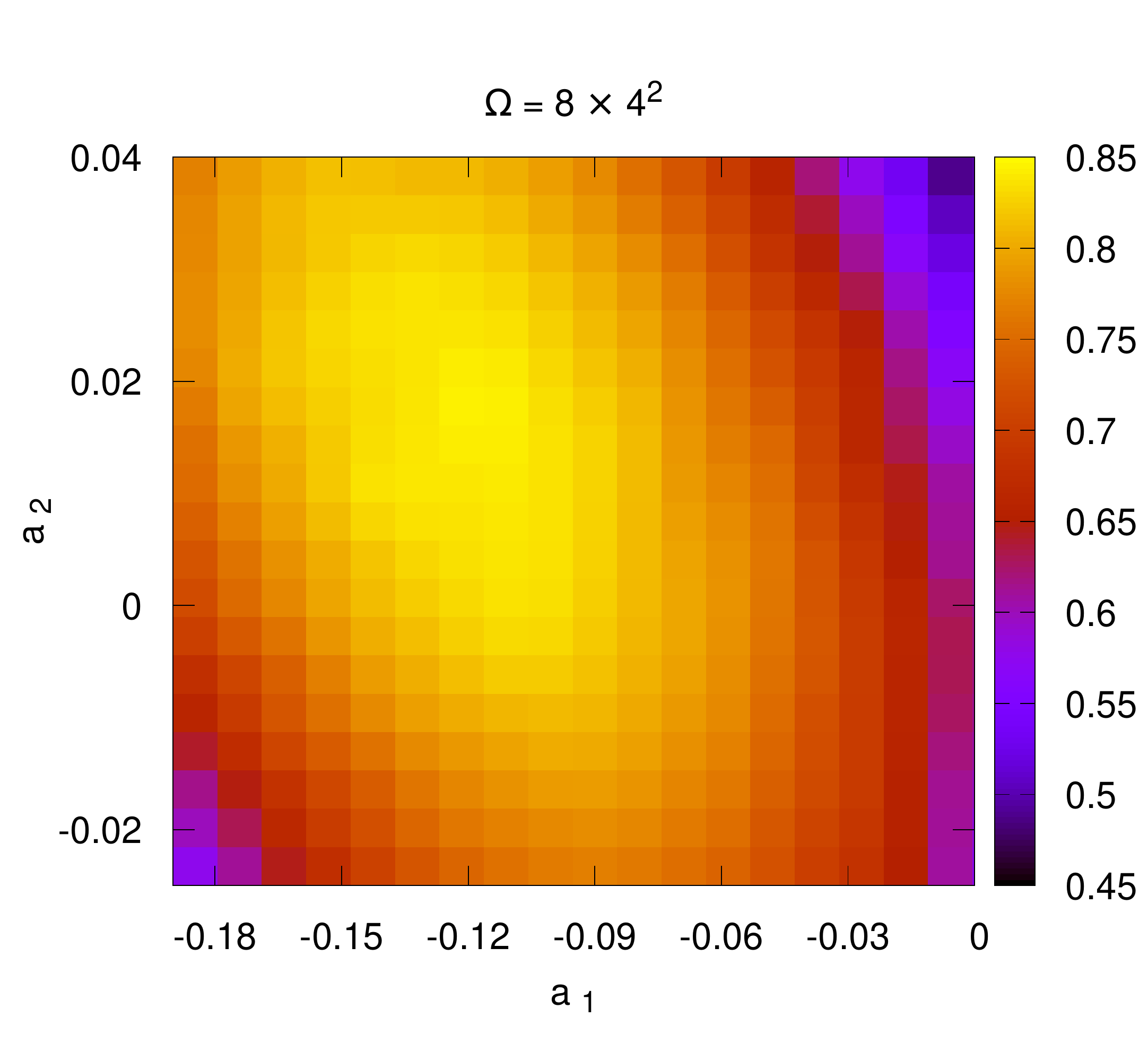} 
\includegraphics[width=0.45\textwidth]{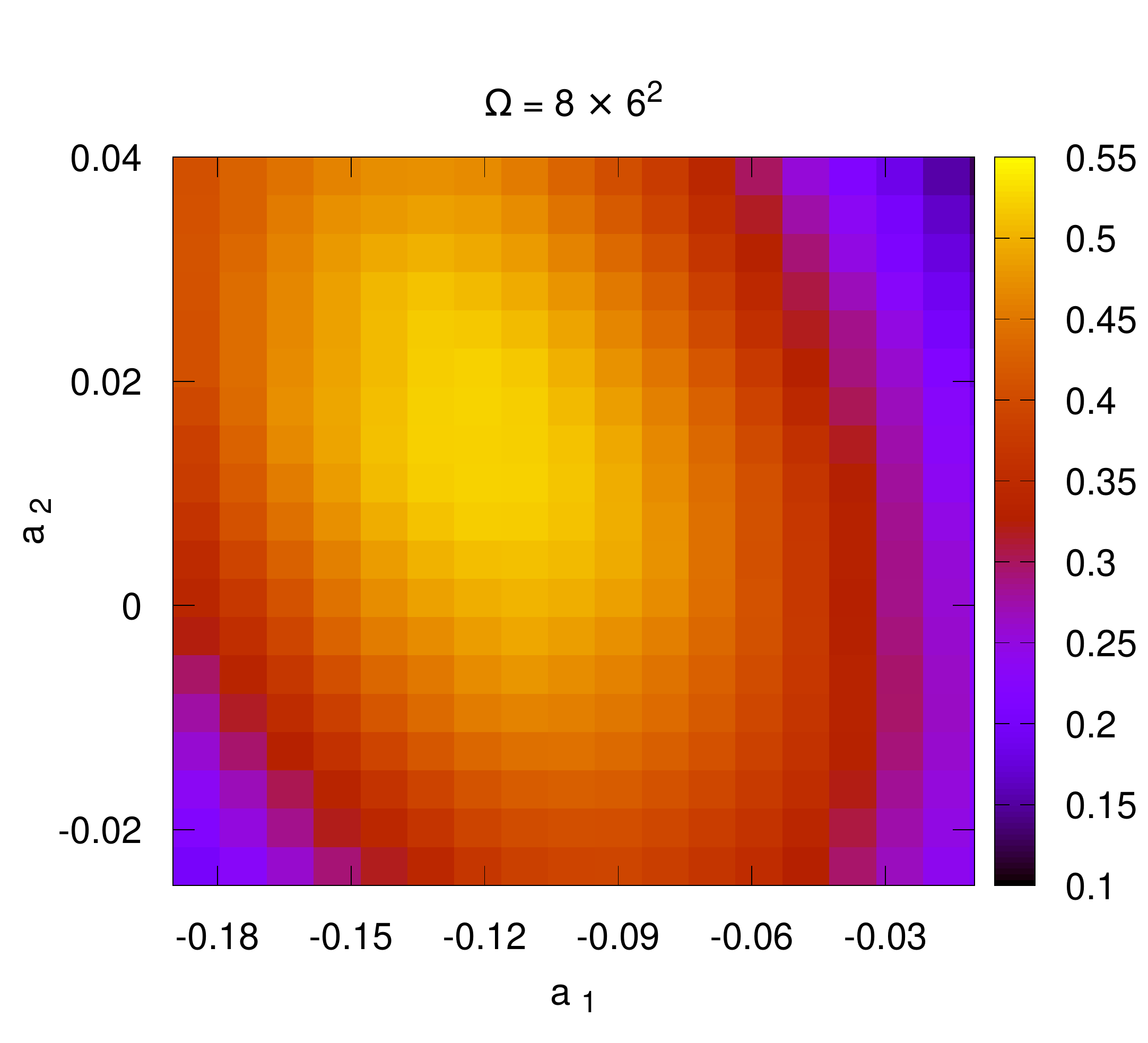}
\includegraphics[width=0.45\textwidth]{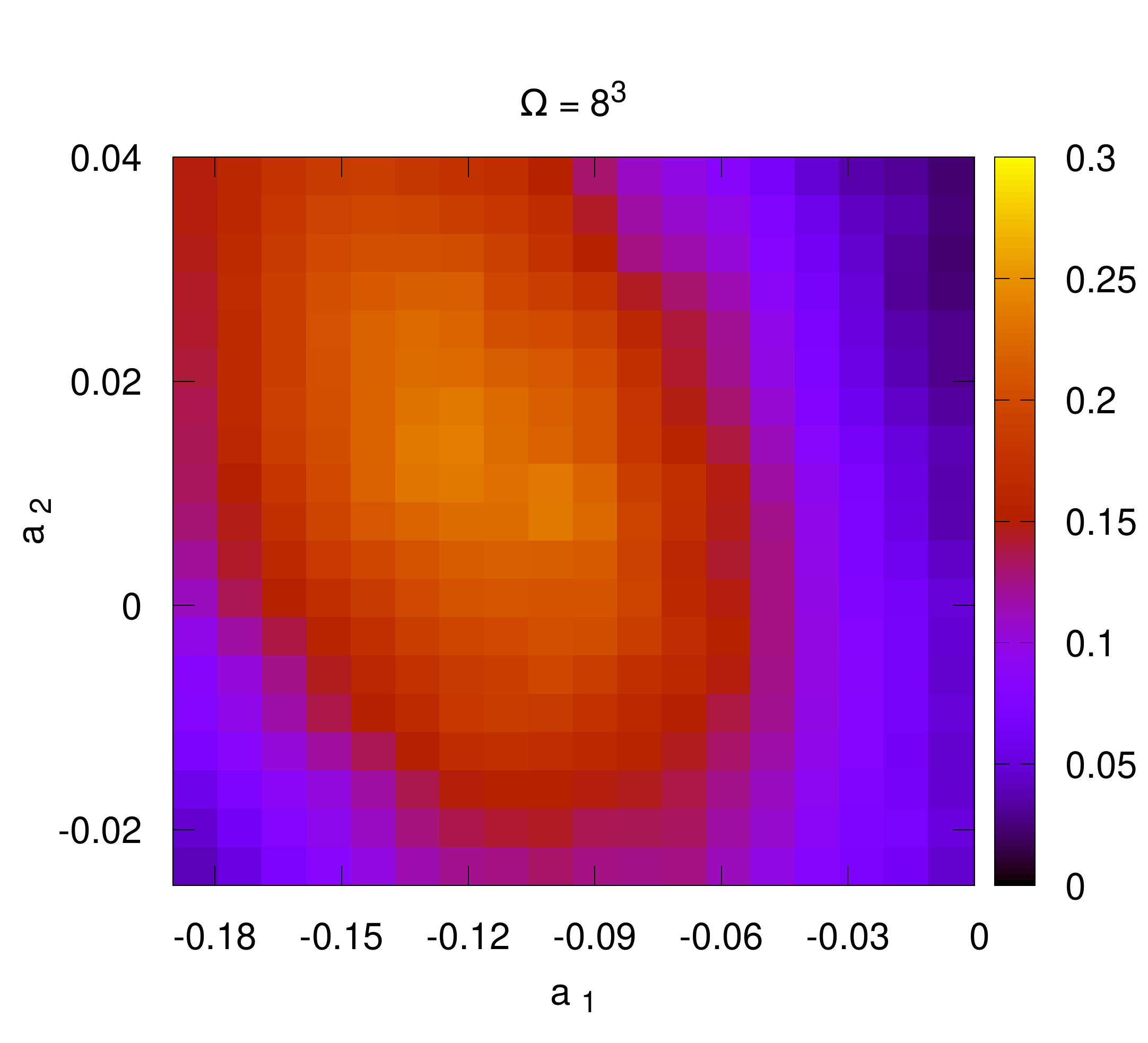}
\includegraphics[width=0.45\textwidth]{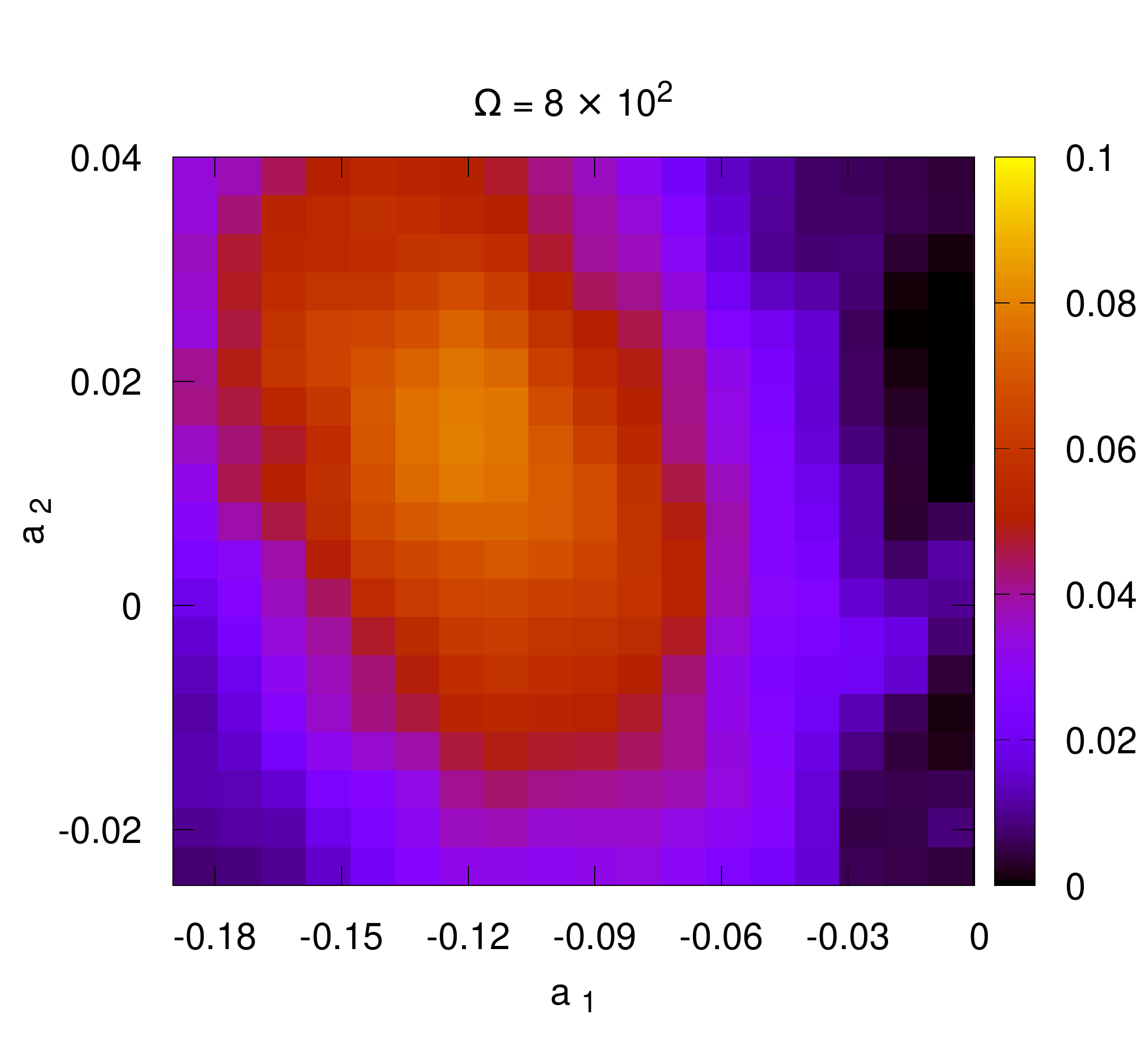}
\caption{Average sign over the space of Fourier coefficients $a_1$ and $a_2$ of Eq.~(\ref{cutparam}) with the $(a_1,a_2)$ setup. Scans were carried out on $\Omega=8\times4^2,8\times6^2,8^3$ and $8\times10^2$ lattices at $\beta=0.4$ and $\mu^2=0.15$.}
\label{fig:a1a2scan}
\end{figure*}
\end{widetext}

Figure~\ref{fig:a1a2scan} shows the average sign over the space of coefficients $a_1$ and $a_2$ of the parametrization described by Eq.~(\ref{cutparam2}) in the $(a_1,a_2)$ setup with $\beta=0.4$ and $\mu^2=0.15$. Each scan was performed at different spatial sizes with temporal size fixed at $N_0=8$.

\clearpage


\end{document}